\documentclass[draftcls,11pt,onecolumn]{IEEEtran}
\usepackage{float}
\usepackage{graphicx}
\usepackage{clrscode}
\usepackage{stfloats}
\usepackage{paralist}
\usepackage{cite}
\usepackage{comment}
\usepackage{graphicx}
\usepackage{amsmath, amssymb, mathrsfs, amsfonts, amsthm}
\usepackage{epsfig, epstopdf}
\usepackage[tight,footnotesize]{subfigure}
\usepackage{algorithm}
\usepackage{color}
\usepackage[usenames,dvipsnames,svgnames,table]{xcolor}
\usepackage{bbold}

\usepackage{chemarrow}

\theoremstyle{theorem}
\newtheorem{theorem}{Theorem}%[section]

\newtheorem{example}{Example}

\theoremstyle{remark}

\def\msKappa{ \mathbf{\zeta}}

\newcommand{\E}{\mathbb{E}}

\newcommand{\Ex}{\mathsf{exp}}

\newcommand{\argmin}{\arg\!\min}              % newest
\newcommand{\argmax}{\arg\!\max}

\begin{document}

\title{Type Based Sign Modulation and its Application for ISI mitigation in Molecular Communication 
}

\author{Reza Mosayebi,
           Amin Gohari,
           Mahtab Mirmohseni,
           and  Masoumeh Nasiri Kenari\\
  			Department of Electrical Engineering, Sharif University of Technology\\
  			%Emails: mosayebi@ee.sharif.edu, \{aminzadeh, mirmohseni, mnasiri\}@sharif.edu.\\
  			}

\maketitle

%eqn:inequ-left-g
%eqn:kjfnb1

\begin{abstract}
An important challenge in design of modulation schemes for molecular communication is positivity of the transmission signal (only a positive concentration of molecules can be released in the environment). This restriction makes handling of the InterSymbol Interference (ISI) a challenge for molecular communication. Previous works have proposed use of chemical reactions to remove molecules from the environment, and to effectively simulate negative signals. However, the differential equation describing a diffusion-reaction process is non-linear. This precludes the possibility of using Fourier transform tools.  In this paper, a solution for simulating negative signals based on the diffusion-reaction channel model is proposed. While the proposed solution does not exploit the full degrees of freedom available for signaling in a diffusion-reaction process, but its end-to-end system is a linear channel and amenable to Fourier transform analysis.  Based on our solution, a modulation scheme and a precoder are introduced and shown to have a significant reduction in error probability compared to previous modulation schemes such as CSK and MCSK. The effects of various imperfections (such as quantization error) on the communication system performance are studied.  
\end{abstract}


\section{Introduction}
% From not submitted paper
%Nano communication is a challenge in the design of nano-scale machines, which are envisioned for their applications in health, biomedicine, industry and, environment 
%engineered systems, etc \cite{AAA}. 

Diffusion-based molecular communication is one of the promising mechanisms for communication among nano-scale machines. In such systems, molecules are used as  information carriers instead of the electromagnetic waves. Information may be encoded in the number, type or release time of the molecules. Released molecules gradually diffuse in the medium, resulting in channel memory and Inter-Symbol Interference (ISI). One of the main theoretical challenges of molecular communication is handling of the channel memory.

While ISI is a common issue in classical communications, it  is more challenging and prominent in the context of molecular communication because one cannot readily combat ISI with classical channel equalization techniques. This is due to the fact that molecular communication systems are \emph{positive systems}, i.e., transmitter can only release a {positive} amount of concentration of a specific molecule into the medium. As a result, consecutive transmissions at the transmitter side can only contribute positively to the molecules at the receiver side, and cannot cancel out the previous transmissions. 

One approach to handle \emph{negative} signals is to add a constant value (dc term) to both negative and positive transmission concentrations to make all the transmitted concentrations positive. For instance, to transmit amplitudes $-10$ and $10$, we can consider a dc term of $10$, i.e., making transmitted concentrations as $0$ and $20$ for negative and positive messages, respectively. However, this will increase the number of transmitted molecules (transmission power) and may not be appropriate. In \cite{Fick}, negative signals are handled by assuming that the input signal is the derivative of molecule concentration. The derivative of a positive signal can be both positive and negative. Thus, when we want to send positive signals, it is suggested to increase the current level of diffusion and if we want to send negative signals, we reduce the current level of transmission. However, the fact that the input signal has to be the derivative of some non-negative waveform does impose restrictions on the input signal. How this constraint translates into the Fourier domain is a challenge in employing the model of \cite{Fick}. 

To simulate negative signals, another known promising idea exploits a unique feature of molecular communication, namely chemical reactions. The general idea is that when molecules of different type bond together, they produce new molecule types and reduce the concentration of original molecule types. In \cite{enzym1}, the release of enzymes throughout the environment to put down the remaining molecular concentration from previous transmissions is proposed. However, this would also weaken the direct link between the transmitter and the receiver and would imply higher transmission powers. Chemical reactions are also employed in design of molecular logic gates like XOR, with two types of molecules that nullify each other \cite{gateXOR}.
Two works that consider chemical reactions to simulate negative signals are \cite{newadd1} and \cite{Gold}. Authors in \cite{newadd1} propose that one puts information on molecules of type $A$, but also follow it by the release molecules of type $B$ (called a poison signal) to chemically cancel effect of remaining ISI terms; thus, molecules of type $B$ are playing the role of negative signals.  In \cite{Gold}, the idea of using $\text{H}^{+}$ and $\text{OH}^{-}$ ions is proposed wherein transmitting each of these ions reduces the concentration of the other one in the medium. Therefore these ions can play roles of positive and negative signals. Diffusion-reaction equations are used to describe the variation of the concentration of ions in the medium. The main challenge of this system is that these diffusion-reaction equations  are \emph{nonlinear}, making its analysis a challenging problem. 

A different suggestion is given in \cite{newadd2} to use two molecule types of $A$ and $B$ as  primary and secondary signals to mitigate ISI. Information is transmitted on the primary signal, via molecules of type $A$. The number of molecules of type $A$ at each time slot are either due to the current transmission, or previous transmissions (ISI term). To convey the ISI term to the receiver, the idea is that after each transmission of primary molecules of type $A$, with a certain delay, a proportional amount of molecules $B$ is emitted to the medium. No reaction occurs between molecules of type $A$ and $B$, and molecules of type $B$ experience a similar channel as molecules of type $A$. Observing that molecules of type $B$ are released after a delay, the idea is to treat the number of arriving molecules of type $B$ as the ISI term of molecules of type $A$ that we wish to cancel. Then, the receiver subtracts the number of received molecules of type $A$ and $B$ to suppress the ISI. A numerical optimization is run to find the best choice of delay and the amount of released molecules of type $B$ to minimize the error probability. While this approach is shown to reduce the ISI, its performance is obtained numerically.

 To sum this up, previous works consider four solutions: (1) adding a dc to make the signal positive, (2) coding information on the signal derivative, (3) using chemical reactions, and (4) using pre-equalization  (with primary and secondary signals). 

Our goal is to find a solution to simulate the effect of negative signals such that the resulting system still \emph{remains linear}. The property of linearity will allow us to apply classical communication approaches, such as spectral transformations, for communication and detection paradigms in the general multiple transmitter and multiple receivers setting. In particular, this scheme will permit us to use a \emph{precoder} filter to compensate the adverse effects of the channel. 

\begin{figure}
\centering
\subfigure[]{
 \centering
\includegraphics[width=0.31\linewidth]{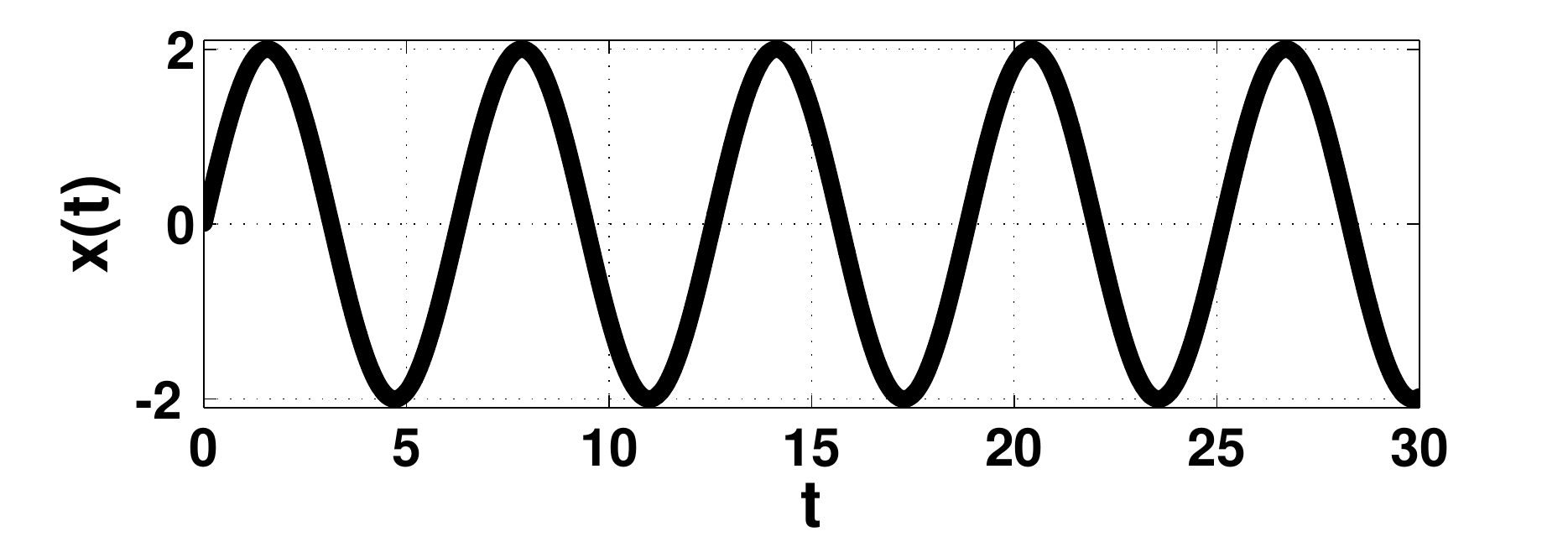}      %%%%%%%%%%%%%%%%%%%% 6 
}
\vspace{-0.1cm}
\subfigure[]{
\centering
\includegraphics[width=0.31\linewidth]{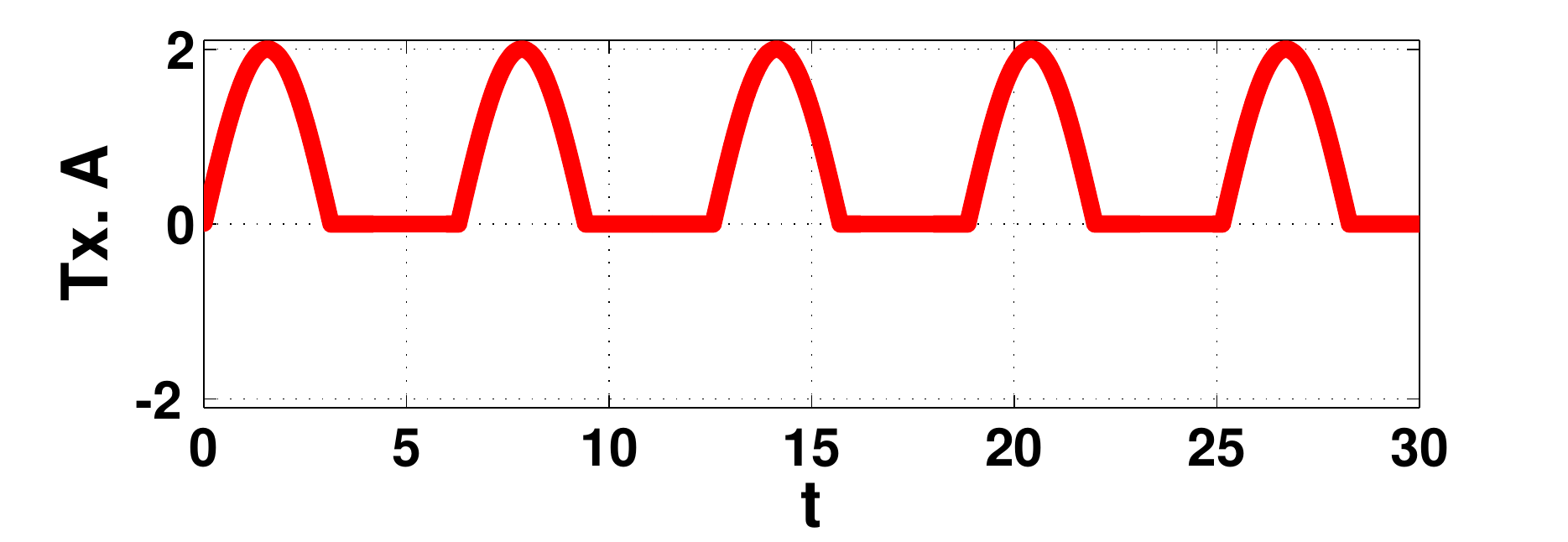}      %%%%%%%%%%%%%%%%%%%% 6 
}
\vspace{-0.1cm}
\subfigure[]{
 \centering
\includegraphics[width=0.31\linewidth]{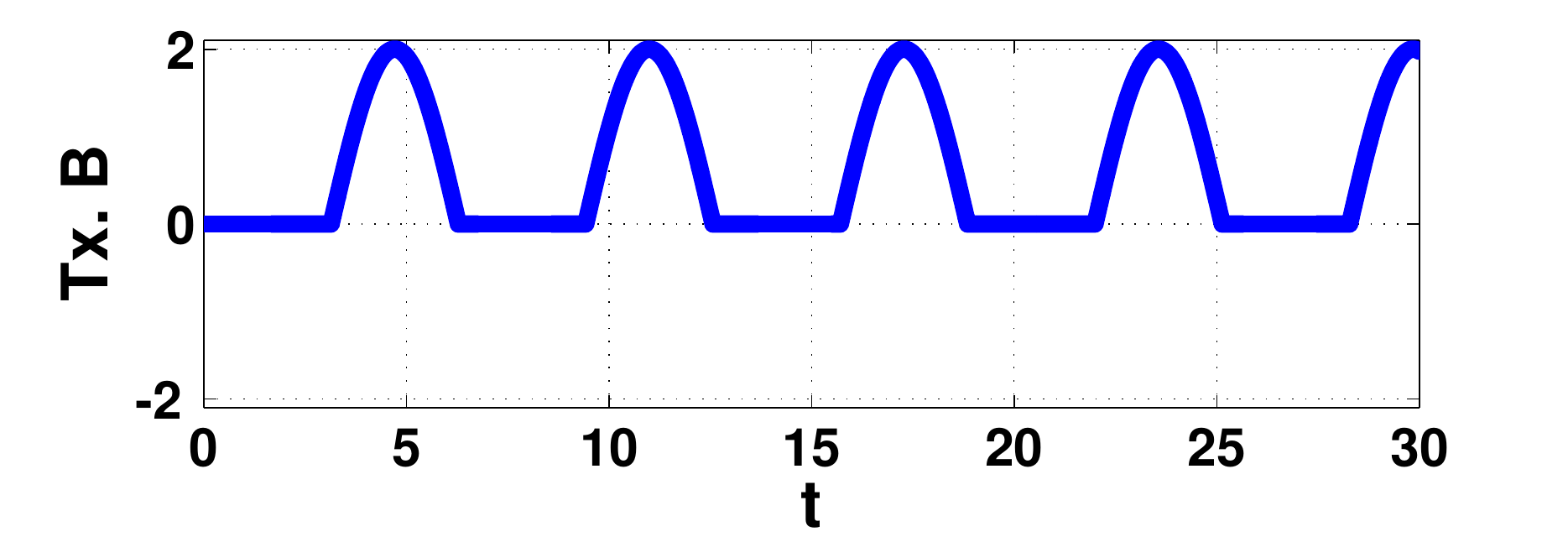}      %%%%%%%%%%%%%%%%%%%% 6 
}
\vspace{-0.1cm}
\caption{Type based sign modulation (a): transmission signal $x(t)=2 \sin(t)$, (b): $|x(t)|\cdot\mathbf{1}[x(t)\geq 0]$ transmitted via molecules of type $A$, (c): $|x(t)|\cdot\mathbf{1}[x(t)\leq 0]$ transmitted via molecules of type $B$.}
\label{fig.Signal}
\vspace{-0.4cm}
\end{figure}

We propose to exploit the fact that any arbitrary real signal is the \emph{difference} of two positive signals. The idea is to use two different types of molecules (denoted by molecules $A$ and $B$) for positive and negative transmissions. In other words, when we would like to send a positive signal, we release molecules of type $A$, and for sending a negative signal, we release molecules of type $B$. That is, if the transmitter wants to transmit signal $x$, if $x>0$ then $x$ moles of molecules of type $A$ will be released, and if $x<0$, then $|x|$ moles of molecules of type $B$ will be inserted into the medium. Fig. \ref{fig.Signal} demonstrates this scheme for transmitting $x(t)= 2 \sin(t)$ into the medium. In this transmission strategy, the difference of the concentration of molecules of types $A$ and $B$ is equal to the signal $x(t)$. We call this signaling the Type based Sign (TS) modulation. 

Our solution has similarities and differences with each of the previous solutions. While our solution is not based on chemical reactions as in \cite{enzym1, Gold}, crucially we see that the resulting system remains linear; in fact, chemical reactions can be utilized for an improved performance. Next, our idea is similar to the one given in \cite{newadd1, newadd2} in the sense that we look at the \emph{difference} of concentrations of molecules of type $A$ and $B$ to identify the signal. But our approach differs from the ones given in \cite{newadd1, newadd2} as we use molecules $A$ and $B$ in a \emph{symmetric way} for positive and negative amplitudes, whereas in \cite{newadd1, newadd2} molecules of type $A$ are the primary and carriers of information, while molecules of type $B$ follow molecules of type $A$ by mirroring them appropriately in order to reduce the channel ISI. Thus, molecules of type $B$ are treated as secondary or poison signals. On the other hand, our approach for canceling the ISI term is via a \emph{precoder} for an effective channel derived using the diffusion-reaction equations. Furthermore, the conceptual reason for taking the {difference} of concentrations of molecules of type $A$ and $B$ is not because we view one as a poison for the other, but because  the difference of concentrations follows a linear differential equation that is not affected by the reaction rate between molecules.

Furthermore, there is a unique feature in our approach. To explain this, let us consider an open unbounded medium with no drift and no reaction. Then, if we release one molar of molecules at the origin at time zero, the concentration of molecules equals
\begin{align}
\rho(t, \vec{r}) = \frac{\mathbf{1}[t>0]}{(4 \pi D t)^{1.5}} \Ex(-\frac{\|r\| _2 ^2 }{4Dt}),
\end{align}
where $\rho(t, \vec{r})$ is the concentration at time $t$ at location $\vec{r}$. For large values of $t$, $\rho(t, \vec{r})$ vanishes like a constant times $t^{-1.5}$, as 
$$\Ex(-\frac{\|r\| _2 ^2 }{4Dt})\approx 1.$$
This inverse polynomial decay is slow and leads to large values of ISI.\footnote{For bounded regions, the decay can be become exponentially fast, like $e^{-ct}$; for instance, see  \cite[p. 3275]{Francesco} for a bounded region with a reflexive boundary condition. However, one should note that the value of $c$ is proportional to the diameter of the region (it depends on the Poincar\'e constant of the medium, which is proportional to its diameter \cite{Bebendorf}). As a result, the decay can be still very slow and lead to large values of ISI. }
On the other hand, we get much faster decays if we allow for chemical reactions. For a broad class of chemical-reaction equations,  it is shown in \cite{Desvillettes2, Desvillettes1} (using the entropy method) that the concentration of molecules converges exponentially fast in time to its stationary distribution like $e^{-t}$.\footnote{Intuitively speaking, the higher the concentration of the reactants, the higher the volume of the reaction. As a result, an exponential drop in concentrations are expected. See also the example in  \cite[p. 3275]{Francesco} which shows that the rate of decay is influenced directly by the chemical reaction rate.} To sum this up, chemical reactions allow for fast exponential decays. This is helpful for combatting the ISI, as the effect of previous transmissions dies out faster. But chemical reactions not only decrease the ISI terms, but also the direct link. In fact, the approach of   \cite{enzym1} based on the release of enzymes has this drawback. However, the unique feature of our approach is that it allows us to overcome this difficulty because the differential equation that governs the difference of molecules of type $A$ and $B$ is not affected by any of chemical reaction rates. Thus, in an intuitive sense, chemical reactions are utilized in a dimension perpendicular to the direction of the signal propagation.

Our contribution can be summarized as follows: (i) we propose a new type based sign (TS) signaling strategy to introduce negative signals for molecular communication (ii) the TS signaling is utilized for precoder design. The effect of various imperfections on the error probability of the precoder are studied; supporting analytical results and numerical simulations are also given.

The remainder of this paper is organized as follows. In Section \ref{sec:system-model}, Fick's law of diffusion as well as reaction-diffusion equations are reviewed.  Section \ref{sec:TSmodul} gives the principles of the TS modulation. The scheme is applied to a time slotted communication setting in Section \ref{sec:tssig}. In Section \ref{sec:model}, we derive a precoder, utilizing the TS modulation. In Section \ref{section:imperfect-reaction}, the effect of imperfect chemical  reactions on the noise variance at the receiver is studied. Section \ref{Numeric} provides extensive numerical results to verify the various claims given in the paper, and finally Section \ref{sec:conc} concludes the paper.
%%  
%%
%%
%%
%%
%%
%%%%%%%%%%%%%%%%
%%%%%%%%%%%%%%%%
%
\section{Preliminaries: Diffusion and Reaction-Diffusion}\label{sec:system-model}
Diffusion of molecules can be studied from either a microscopic or  macroscopic perspectives, depending on the number of released molecules. If the number of released  molecules is small,  the microscopic perspective is relevant and the movement of individual molecules should be studied through Brownian motion and stochastic differential equations. If possibility of chemical reaction between molecules of different types is considered, one will have Brownian motion with interacting particles. On the other hand, if the number of released  molecules is large, the macroscopic perspective is relevant and the variation of the concentration of molecules should be studied through the deterministic Fick's laws of diffusion. If chemical reaction between molecules also exists, one should look at reaction-diffusion differential equations (still a deterministic equation). Herein, we assume that the number of released molecules is large and utilize the macroscopic perspective. 

The macroscopic diffusion of a molecules over an medium is described as follows: assume that the medium has a time-invariant flow velocity  $v(\vec{r})$ at location $\vec{r}$.  The Fick's second law of diffusion states that
\begin{align}
\frac{\partial \rho(t,\vec{r})}{\partial t} &= D \nabla ^2 \rho(t,\vec{r}) -  \nabla\cdot  (v(\vec{r}) \rho(t,\vec{r}))+s(t,\vec{r}),
\label{eq01} 
\end{align}
where  $\rho(t,\vec{r})$ is the concentration of molecule at time $t$ and position $\vec{r}$, $D$ is the diffusion coefficients of molecules, and $s(t,\vec{r})$ denotes the concentration of molecule production rate at point $\vec{r}$ (by the transmitter), \emph{i.e.,} the number of molecules added to the environment between time $[t, t+dt]$ in a small volume  $dV$ around $\vec{r}$ is equal to $s(t,\vec{r})dVdt$. Equation \eqref{eq01} can be understood as follows: any change in $\rho(t,\vec{r})$ is due to the following factors:
\begin{itemize}
\item The term $D\nabla ^2 \rho(t,\vec{r})$ due to \emph{medium inhomogeneity}: molecules flow from area of high concentration to areas of low concentration, and the rate of this flow is proportional to the difference of concentrations. Thus, the gradient of $\rho(t,\vec{r})$ gives the flow due to medium inhomogeneity. But the mere existence of a flow does not necessarily imply a change in concentration of molecules at $\vec{r}$. For instance, in a uniform flow, the number of molecules that enter $\vec{r}$ equals the number of molecules that exit  $\vec{r}$. Concentration of molecules at $\vec{r}$ changes if there is a mismatch between the incoming  and outgoing flows  at $\vec{r}$, \emph{i.e.,} variations in the gradient of $\rho(t,\vec{r})$, which is  $D\nabla ^2 \rho(t,\vec{r})$.
\item The term  $\nabla\cdot  (v(\vec{r}) \rho(t,\vec{r}))$ is due to the velocity of the medium. The velocity of the medium can alter concentration of molecules at $\vec{r}$, but as velocity mixes molecules from a neighborhood of $\vec{r}$, it should be considered in conjunction with the concentration of molecules in the vicinity of $\vec{r}$. This is reflected in computing the divergence of the product $v(\vec{r}) \rho(t,\vec{r})$.
\item The molecule production rate $s(t,\vec{r})$ at $\vec{r}$ directly contributes to a change in $\rho(t,\vec{r})$. The molecule production rate is due to the transmitter(s).
\end{itemize}
Let us denote the medium by a bounded region $\Omega\subset \mathbb{R}^n$, and its boundary by $\partial\Omega$. To ensure that \eqref{eq01} has a unique solution for $\rho(t,\vec{r})$, initial, boundary conditions as well as some technical mathematical assumptions about the boundary of $\partial\Omega$ and $s(t,\vec{r})$ will be imposed \cite[p. 16--17]{Pao}. Throughout this paper, we assume \emph{initial rest and zero boundary condition}. The medium is in initial rest if it is zero and remains zero until a non-zero input is applied to the medium by the transmitter(s), \emph{i.e.,} if the molecule production rate $s(t,\vec{r})=0$ for all $t\leq t_0$ and becomes non-zero for some $t>t_0$, we assume that 
$\rho(t_0,\vec{r})=0$ for all $\vec{r}$. 

 Our TS signaling strategy can be utilized with any type of boundary condition as long as a zero boundary condition is assumed, \emph{e.g.} for a Dirichlet boundary condition, a perfectly  absorbing boundary is assumed; for a Neumann boundary condition, a perfectly reflective boundary is assumed.\footnote{We  assume that the boundary of the region $\partial\Omega$ has the desired smoothness properties, \emph{e.g.,} for the Dirichlet boundary condition, we assume that it has the outside strong sphere property (for every point $\vec{r}\in \partial\Omega$ on the boundary, one can find a closed ball $B$ outside $\Omega$ such that $B\bigcap \partial\Omega=\{\vec{r}\}$).} Finally, the molecule production rate $s(t,\vec{r})$ and velocity $v$ are assumed to be  H\"older continuous \cite[p. 16--17]{Pao}.\footnote{A function $f$ is said to be  H\"older continuous if  there are non-negative real constants $C$, $\alpha$, such that
$$|f(x) - f(y)| \leq C\|x-y\|^{\alpha}$$
for all $x$ and $y$ in the domain of $f$, and $\|\cdot\|$ is any arbitrary norm (all norms are equivalent in a finite dimensional space).} 
 Under these conditions,  \eqref{eq01} is guaranteed to have a unique solution  $\rho(t,\vec{r}))$ that is continuously differentiable in $t$ and twice continuously differentiable in $\vec{r}$ \cite[p. 16--17]{Pao}.

A zero boundary condition in conjunction with initial rest imply that the differential equation \eqref{eq01} describes a linear time invariant (LTI) system when we view $s(t,\vec{r})$ as the input and $\rho(t,\vec{r})$ as the output.

\textbf{Diffusion of two  types of molecules:} 
Below, we provide the diffusion equations for two molecule types of $A$ and $B$, in the absence or presence of chemical reactions. 

\emph{The case of no chemical reaction:} Consider diffusion of molecules of type $A$ and $B$ over an environment with flow velocity $v(\vec{r})$.  The Fick's second law of diffusion states that
\begin{align}
\frac{\partial \rho_{A}(t,\vec{r})}{\partial t} &= D_A \nabla ^2 \rho_{A}(t,\vec{r}) -  \nabla\cdot  (v(\vec{r}) \rho_A(t,\vec{r}))+s_{A}(t,\vec{r}),
\label{eq1} \\
\frac{\partial \rho_{B}(t,\vec{r})}{\partial t} &= D_B \nabla ^2 \rho_{B}(t,\vec{r}) -  \nabla\cdot  (v(\vec{r}) \rho_B(t,\vec{r}))+s_{B}(t,\vec{r}),
\label{eq2}
\end{align}
where  $\rho_A(t,\vec{r})$ and $\rho_B(t,\vec{r})$ are the concentration of molecules $A$ and $B$, $D_A$ and $D_B$ are the diffusion coefficients, and $s_A(t,\vec{r})$ and $s_B(t,\vec{r})$ are the  molecule production rates. Observe that equations \eqref{eq1} and \eqref{eq2} are decoupled. 

\emph{Diffusion with chemical reaction:} Assume that molecules of $A$ and $B$ make chemical reactions due to the following equation
\begin{align}\label{eqn:chemical-reaction}
%\ch{A + B <>[ $k_1$ ][ $k_2$ ] Outputs }
A+B \xrightarrow{\msKappa} Outputs
\end{align}
in which $\msKappa$ is the rate of reaction. Note that in \eqref{eqn:chemical-reaction},  the coefficients of $A$ and $B$ are assumed to be equal; hence equal concentrations of molecules $A$ and $B$ cancel out each other. Assuming that the output products of \eqref{eqn:chemical-reaction} do not include either of molecules of type $A$ or $B$, the reaction-diffusion law can be expressed as
\begin{align}
\frac{\partial \rho_{A}}{\partial t} &= D_A \nabla ^2 \rho_{A} -  \nabla\cdot  (v \rho_A)
-\msKappa\rho_A \rho_B+s_{A},\label{eq1av} \\
\frac{\partial \rho_{B}}{\partial t} &= D_B \nabla ^2 \rho_{B} -  \nabla\cdot  (v \rho_B)
-\msKappa\rho_A \rho_B+s_{B}. \label{eq2av}
\end{align}
The new term $\msKappa\rho_A\rho_B$ indicates that when both $\rho_A$ and $\rho_B$ are positive, molecules of type $A$ and $B$ disappear from the environment at a rate $\msKappa\rho_A\rho_B$. Indeed, the larger the values of $\rho_A$ and $\rho_B$, the larger $\msKappa\rho_A\rho_B$, and the faster they die out in the medium (this is like a negative feedback that pulls down  $\rho_A$ and $\rho_B$ once they become positive). Despite its clear intuitive meaning, unfortunately the product term $\msKappa\rho_A\rho_B$ is a  \emph{non-linear} term. In Appendix \ref{somecomment}, we comment on the reaction diffusion equations given in \eqref{eq1av} and \eqref{eq2av} and the effect of the non-linear term on it. Due to the non-linear term, no explicit analytical solution is known for \eqref{eq1av} and \eqref{eq2av} in general. Showing that the above differential equations with proper boundary conditions have a \emph{unique non-negative} solution is non-trivial from a mathematical point of view (\emph{e.g.} see the first two chapters of \cite{Pao}).

\section{Principles of Type Based Sign Modulation}\label{sec:TSmodul}
We assume that the transmitter's action is the molecule production rate $s_A(t,\vec{r})$ and $s_B(t,\vec{r})$ of molecules of type $A$ and $B$ at origin. The functions $s_A(t,\vec{r})$ and $s_B(t,\vec{r})$ are assumed to be H\"older continuous. 

As mentioned in the introduction, in the TS modulation, we encode information in the difference of concentrations of molecules $A$ and $B$. In other words, we only track the difference $\rho_{A}(t,\vec{r})-\rho_B(t,\vec{r})$ as a function of time $t$ and location $\vec{r}$. In particular, at the receiver side, concentrations of both molecules $A$ and $B$ are measured and their difference is considered as the received signal. That is if $\alpha_1$ moles molecules of $A$ and $\alpha_2$ moles molecules of $B$ are measured at the receiver, receiver will interpret the value of $\alpha_1-\alpha_2$ as the received signal. Similarly, if the transmitter injects $\beta_1$ moles molecules of $A$ and $\beta_2$ moles of $B$ in the environment, we interpret it as injecting $\beta_1-\beta_2$. As a result, we need to find the differential equation that describes $\rho_{A}(t,\vec{r})-\rho_B(t,\vec{r})$ in terms of molecule injection difference $s_{A}(t,\vec{r})-s_B(t,\vec{r})$.

\emph{Diffusion with no chemical reaction:} Let us assume that $D_A=D_B=D$. By subtracting (\ref{eq2}) from (\ref{eq1}) we have
\begin{align}
\frac{\partial (\rho_{A}-\rho_{B}) }{\partial t} = D \nabla ^2 (\rho_{A} - \rho_{B}) -  \nabla\cdot  (v(\vec{r}) (\rho_A-\rho_B))
+s_{A}-s_{B}.\label{diff}
\end{align}
Thus, if we define
\begin{align}\rho(t,\vec{r})&:=\rho_{A}(t,\vec{r})-\rho_{B}(t,\vec{r}),\\
s(t,\vec{r})&:=s_{A}(t,\vec{r})-s_{B}(t,\vec{r}),
\end{align}
the $\rho$  follows the same \emph{linear} differential equation, similar to  (\ref{eq2}) and  (\ref{eq1}), when we interpret $s$ as the molecule production rate. Assumption of initial rest and zero boundary condition for $\rho_A$ and $\rho_B$ implies the same condition for their difference.

\emph{Diffusion with chemical reaction:} Assume that molecules of $A$ and $B$ make chemical reactions according to \eqref{eqn:chemical-reaction}, and follow the reaction-diffusion equations of \eqref{eq1av} and \eqref{eq2av} with $D_A=D_B=D$. Observe that by subtracting equations \eqref{eq1av} and \eqref{eq2av}, the non-linear terms $\rho_A\rho_B$ cancel out and we get back equation 
\eqref{diff} (the equation for the no-reaction case).\footnote{The property needed is that the  non-linear terms of equations \eqref{eq1av} and \eqref{eq2av} be the same, so that it vanishes when we subtract the two equations. This can occur for reactions other than \eqref{eqn:chemical-reaction}, \emph{e.g.,} for a reaction of the form
 \begin{align}\label{eqn:chemical-reaction2}
%\ch{A + B <>[ $k_1$ ][ $k_2$ ] Outputs }
2A+3B \xrightarrow{\msKappa} B+Outputs
\end{align}
where ``Outputs" do not involve any molecules of type $A$ or $B$.}

\emph{Benefits and drawbacks of TS:} the benefit of the TS scheme is having allowed for negative input signal while keeping the system linear. This allows for combating ISI, as discussed below. Drawbacks include the use of two molecule types and the loss of degrees of freedom due to the fact that information is encoded only in the difference of molecule densities. Nonetheless, we will see that the TS scheme already outperforms previously proposed schemes using two molecule types.

\emph{Observation noise at the receiver}: while the macroscopic  diffusion process is deterministic, ligand receivers have signal dependent noise, proportional to the number of measured concentration. As the concentration of molecules increases, the observation noise (known as the \emph{particle counting noise} \cite{Akan,noise}) increases. More specifically, if the receiver is located at $\vec{r}^*$, and we denote the concentration of molecule type $A$ around the receiver by $\rho_{A}(t, \vec{r}^*)$, the receiver's measurement of molecules of type $A$ at time $t_i$ can be written as 
$\rho_A(t_i, \vec{r}^*)+N(t_i)$ where $N(t_i)$ is an additive sampling noise, distributed according to
$\mathcal{N}(0, \frac{1}{V_R}\rho_A(t_i, \vec{r}^*))$ where $V_R$ is the receiver volume \cite{Akan}. We adopt the assumption that the measurement noise at different time instances $t_i$ and $t_j$ are independent conditioned on knowing $\rho(t_i, \vec{r}^*)$ and $\rho(t_j, \vec{r}^*)$.\footnote{As noted in \cite{OurPaperReview}, this assumption is not necessarily true if $|t_i-t_j|$ is very small.}

The above discussion means that it is possible to have small \emph{difference} in concentrations of molecules $A$ and $B$ at the receiver, however each one may have large concentrations; this would degrade the detector performance when we measure the concentration of the individual molecules separately (we incur two separate particle counting noises for measuring molecules of type $A$ and $B$). To put down this noise, one can utilize one of the following two ideas:
\begin{itemize}
\item Use molecules of type $A$ and $B$ that react with each other  according to \eqref{eqn:chemical-reaction}. The reaction puts down the concentration of both molecules, while keeping the difference of the two concentrations intact. 
\item We may use a molecules of type $C$ at the receptor area which together with molecules of $A$ and $B$ make a reaction and disappear. That is, the receiver releases molecules of type $C$ in its vicinity just before making the measurements. Molecules of type of $C$ are such that the following chemical reaction holds:
 \begin{align} \label{reaction}
%1A + 1B +\gamma C   \ch{<>[ $k_1$ ][ $k_2$ ]} Outputs
1A + 1B +\gamma C \xrightarrow{\msKappa} Outputs
 \end{align}
But that $A$ or $B$ alone with $C$ do not form any reaction. Note that the chemical reaction  is occurring only locally around the receiver. Further assume that the output products of \eqref{reaction} do not include either of molecules of type $A$ or $B$. 
 This would imply that release of $C$ causes the concentration of both $A$ and $B$ to drop by the same amount; but their concentration difference will be preserved, as one can observe from the following
\begin{align}
\frac{\partial \rho_{A}}{\partial t} &= D_A \nabla ^2 \rho_{A} -  \nabla\cdot  (v \rho_A)
-\msKappa\rho_A \rho_B\rho_C^\gamma+s_{A}, \label{eq1avN} \\
\frac{\partial \rho_{B}}{\partial t} &= D_B \nabla ^2 \rho_{B} -  \nabla\cdot  (v \rho_B)
-\msKappa\rho_A \rho_B\rho_C^\gamma+s_{B}. \label{eq2avN}
\end{align}

\begin{example}\label{example1n}
An example is the following reaction (for instance see \cite[Ch.17]{refReaction} ).
\begin{align} \label{Acetone}
   &CH_3COCH_3(aq)+Br_2(aq) \xrightarrow{H^{+}(aq)}  CH_3COCH_2Br(aq) +H^{+}(aq)+Br^{-}(aq)
\end{align}
in which molecules acetone ($CH_3COCH_3$), bromine ($Br_2$) and hydrogen ion ($H^{+}$) play role of molecules A, B and C in (\ref{reaction}), respectively. The diffusion coefficient of acetone and bromine are equal to $1.16 \times 10^{-5} \text{cm}^2/\text{s}$ and $1.18 \times 10^{-5} \text{cm}^2/\text{s}$, respectively (\cite[p. 127]{Diffuse1}). Observe that the diffusion coefficients are not exactly the same, but simulation results provided later show that diffusion coefficient mismatch of up to 10 percent does not significantly degrade the error performance. 
\end{example}

\end{itemize}

\section{TS signalling for a time slotted communication setting}\label{sec:tssig}
In this section, we explicitly work out equations of TS signaling for a time slotted communication in which duration of each slot is $T_s$. Transmission and reception occur at the beginning and end of each time slot respectively. This model is employed in the next section for  precoder design. See  Appendix \ref{somecomment} for a discussion on the tradeoff between $T_s$, diffusion coefficients $D_A$, $D_B$ and reaction rate $\msKappa$ for the reaction diffusion equations given in \eqref{eq1av} and \eqref{eq2av}.

\emph{Encoder:} The encoder has a message $M$ consisting of a sequence of information bits. It encodes this information bits into channel inputs, which is a sequence of transmission levels ${X_k}\in\mathbb{R}$ for the $k$-th time slot. Let us fix a given transmission waveform $w(t,\vec{r})$, which we assume is H\"older continuous and $w(t,\vec{r})=0$ for $t<0$. We also expect $w(t,\vec{r})$ to be zero only when $\vec{r}$ is not in the vicinity of origin $\vec{0}$, which is where the transmitter is located. Then, if $X_k>0$, we produce a molecule production rate of $X_kw(t-kT_s,\vec{r})$ for molecules of type $A$. On the other hand, if $X_k<0$, we produce a molecule production rate of $-X_kw(t-kT_s,\vec{r})$ for molecules of type $B$.
 Then, the transmission densities (of equations \eqref{eq1}, \eqref{eq2} , \eqref{eq1av}, \eqref{eq2av}) are equal to
$$s_{A}(t,\vec{r}) = \sum_{k: X_k\geq 0}X_k\cdot w(t-kT_s,\vec{r}),$$
$$s_{B}(t,\vec{r}) = \sum_{k: X_k\leq  0}(-X_k)\cdot w(t-kT_s,\vec{r}).$$
The transmitted \emph{differential  concentration } (i.e., the difference of concentration of molecules of type $A$ and $B$) equals
$$s(t, \vec{r})=\sum_{k}X_k\cdot w(t-kT_s,\vec{r}).$$

\begin{figure}
\centering 
\includegraphics[width=0.7\textwidth]{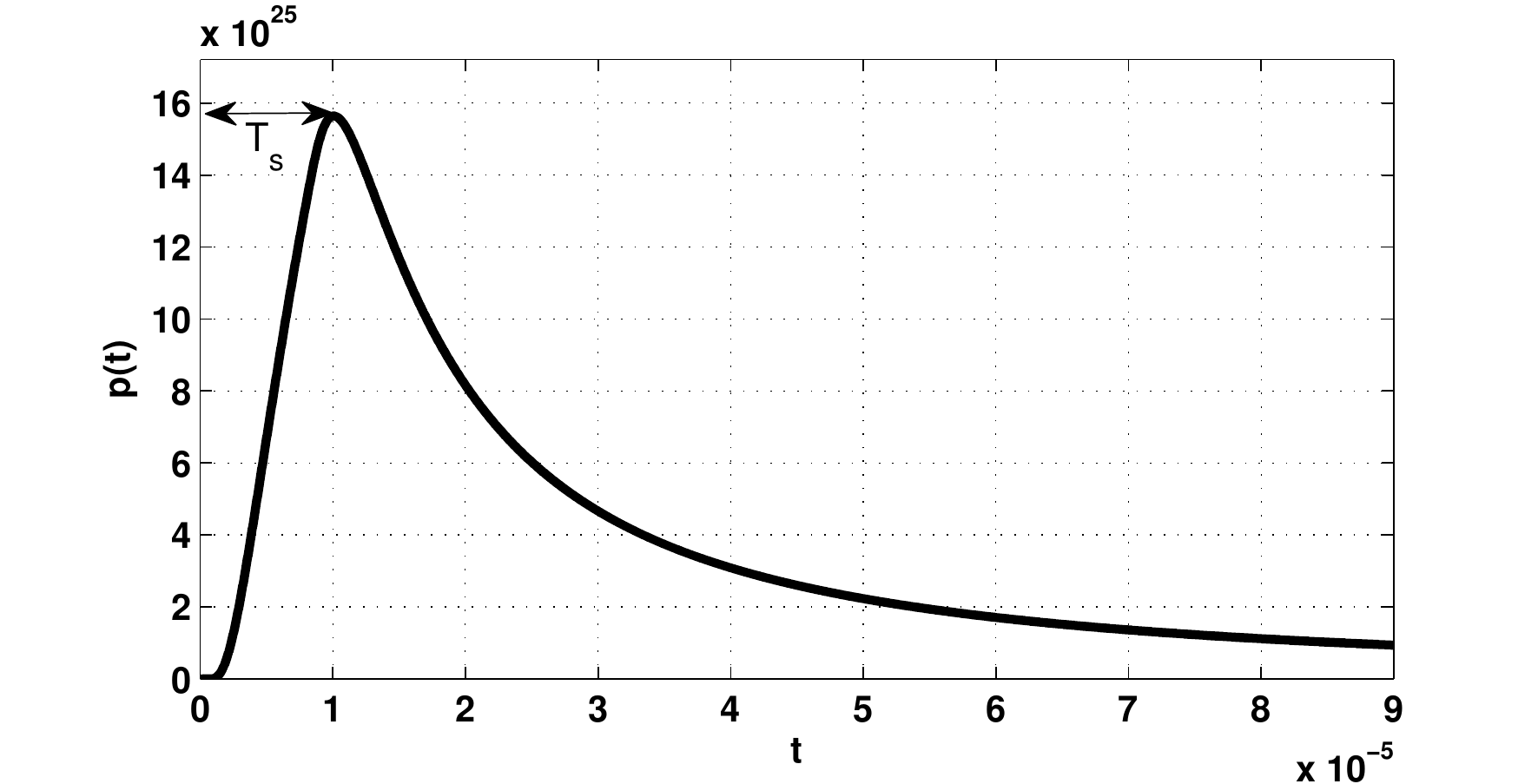}
\vspace{-0.5cm}
\caption{Concentration of molecules at the receiver side ($p(t)$) for transmission of a rectangular pulse. Parameters are    $D= 2.2\times 10^{-9} m^2/s, \| r^{*}\| = 2.15\times 10^{-7}m$, and $T_s = 10 \mu s$. }
\label{fig.channel} %% label for entire figure
\vspace{-0.5cm}
\end{figure}

\emph{Channel:}
 The released molecules  propagate through the medium, possibly react with each other, and will be received at the receiver space, complying with the diffusion laws. That is molecular concentration varies according to both distance between transmitter and receiver and the time receiver is measuring the concentration. Regardless of whether molecules $A$ and $B$ react with each other as they propagate, or whether they react with molecules $C$ at the receiver side, the system that takes in the differential  concentration $s(t, \vec{r})$ and outputs the differential concentration at the receiver $\rho(t, \vec{r}^*)$is linear.  Here $\vec{r}^*$ is the location of the receiver. If the characteristics of the medium does not change, this system is also time-invariant, and hence LTI. The  response of this system to input impulse function $\delta(\vec{r})\delta(t)$ in a 3-dimensional medium without any reaction and drift velocity is given by the green function
\begin{align}
h(t, \vec{r}) = \frac{\mathbf{1}[t>0]}{(4 \pi D t)^{1.5}} \Ex(-\frac{\|r\| _2 ^2 }{4Dt}),
\end{align}
where $D=D_A=D_B$ is the diffusion coefficient of the transmitted molecules and $\| .\|_2$ denotes the $\ell_2$ norm. The differential molecule concentration around the receiver is equal to  $$\rho(t, \vec{r}^*) = \left(\left(\sum_{k}X_k\cdot w(t-kT_s, \vec{r})\right)\star h(t,\vec{r})\right)\big|_{(t, \vec{r}^*)}$$ where $\star$ is the convolution operator. Then,
$$\rho(t, \vec{r}^*) =\sum_{k}X_k\cdot p(t-kT_s),$$
where $$p(t)=\left(w(t, \vec{r})\star h(t,\vec{r})\right)\big|_{(t, \vec{r}^*)}.$$
A typical sample of $p(t)$ is depicted in Fig.~\ref{fig.channel} for a smooth pulse-like $w(t,\vec{r})$ waveform. The actual shape of $p(t)$ depends on our choice of $w(t,\vec{r})$, but generally speaking the intensity of $p(t)$ initially increases and then decreases. It is suggested in  \cite{Akan} to take the sampling time interval $T_s$, at the time instance where the peak of $p(t)$ occurs, i.e.,
\begin{align}T_s=\argmax_{t}p(t)\label{eqn:defTs}.\end{align}
For example, the value of $T_s$  in Fig.~\ref{fig.channel} is $1\times 10^{-5}$.
 With this choice and letting \begin{align}p_j = p(jT_s),\end{align} we have
 \begin{align}\label{condition}
 p_0> p_1 > \cdots > p_L >0
 \end{align}
The value of $L$ is taken large enough that we may practically assume that $p(kT_s) = 0 ~\forall k>L$. Interestingly, a new reason for making this choice of $T_s$ arises in Section \ref{sec:stability}.

\emph{Receiver:} The receiver measures the molecule concentration at time instances $kT_s$ at location $\vec{r}^*$, denoted by $Y_j = Y(jT_s)$ which can be expressed as $\rho(jT_s, \vec{r}^*)+N_j$ where $N_j$ is the  measurement noise. The  measurement noise can be calculated as follows: assume that at time $jT_s$, the concentration of molecules of type $A$ and $B$ are denoted by $\rho_A(jT_s, \vec{r}^*)$ and $\rho_B(jT_s, \vec{r}^*)$, we know that their difference is $\rho_A(jT_s, \vec{r}^*)-\rho_B(jT_s, \vec{r}^*)=\rho(jT_s, \vec{r}^*)$; the actual values of $\rho_A(jT_s, \vec{r}^*)$ and $\rho_B(jT_s, \vec{r}^*)$ depend on whether the involving molecules react and cancel out each other. The measurement counting noise of molecules of type $A$ and $B$ are independent conditioned on $\rho_A(jT_s, \vec{r}^*)$ and $\rho_B(jT_s, \vec{r}^*)$, and distributed according to
\begin{align}
N^A_j &\sim \mathcal{N}\left(0, \frac{1}{V_R}\rho_A(jT_s, \vec{r}^*)\right), \\
N^B_j &\sim \mathcal{N}\left(0, \frac{1}{V_R}\rho_B(jT_s, \vec{r}^*)\right),
\end{align} 
where $V_R$ is the receiver volume. 
Therefore, the total counting noise $N_j$ conditioned on $\rho_A(jT_s, \vec{r}^*)$ and $\rho_B(jT_s, \vec{r}^*)$ is distributed as
\begin{align}
N_j \sim \mathcal{N}\left(0, \frac{1}{V_R}(\rho_A(jT_s, \vec{r}^*)+\rho_B(jT_s, \vec{r}^*))\right).\label{eqn:noisevareq1}
\end{align} 
Furthermore, the counting noise $N_j$ in different time-slots (for different values of $j$) are assumed to be independent of each other, when conditioned on the densities of $A$ and $B$ in the time slots. 

Equation \eqref{eqn:noisevareq1} is not in an explicit form. Thus, we first consider the case of fast reactions (perfect reactions), for which an explicit formula can be found. Imperfect reaction is considered in Section \ref{section:imperfect-reaction}. 

Assume a fast reaction, that is, the reaction-diffusion equations given in \eqref{eq1av} and \eqref{eq2av} or the one given in \eqref{reaction}  with large reaction rate $\msKappa$. Fast reactions have been the subject of many studies in the literature, \emph{e.g.,} see \cite{Fast1, Fast2, Fast3, Fast4}. A typical result, shown in different papers for different settings, is that the solutions of the fast reaction differential equations converge as $\msKappa\rightarrow\infty$. Furthermore, they converge to the solution of the \emph{instantaneous reaction}. Consider the reaction-diffusion equations given in \eqref{eq1av} and \eqref{eq2av}. As discussed in \cite{Fast1}, as we increase $\msKappa$, we observe a spatial segregation of areas with at which molecules of type $A$ and $B$ are present; and the reaction mainly happens in the vicinity of the interface of regions that separate molecules of type $A$ and $B$. Similarly,  with the use of molecule $C$ as in \eqref{reaction}, a similar phenomenon occurs in the vicinity of the receiver. Thus, if $\rho(jT_s, \vec{r}^*)>0$, there are more molecules of type $A$ than $B$, molecules of type $B$ get canceled out and, $\rho_A(jT_s, \vec{r}^*)=\rho(jT_s, \vec{r}^*)$ and $\rho_B(jT_s, \vec{r}^*)=0$. If
$\rho(jT_s, \vec{r}^*)\leq 0$ after the reaction, $\rho_A(jT_s, \vec{r}^*)=0$ and $\rho_B(jT_s, \vec{r}^*)=|\rho(jT_s, \vec{r}^*)|$. Hence, in both cases
$$\rho_A(jT_s, \vec{r}^*)+\rho_B(jT_s, \vec{r}^*)=|\rho(jT_s, \vec{r}^*)|.$$
Thus, we can write
\begin{align}\label{noprecoder-1}
Y_j = \sum_{k=0}^{L} p_k X_{j-k} +N_j,
\end{align} 
where the noise $N_j$  is distributed according to
\begin{align}\label{noprecoder-12}
N_j \sim \mathcal{N}\left(0, \frac{1}{V_R}\left|\sum_{k=0}^{L} p_k X_{j-k}\right|\right).
\end{align} 
Thus, we find that  the probability density function (pdf) of $Y_j$ is given by
\begin{align}\label{noprecoder}
Y_j \sim \mathcal{N}\left(\sum_{k=0}^{L} p_k X_{j-k}, \frac{1}{V_R}\left|\sum_{k=0}^{L} p_k X_{j-k}\right|\right).
\end{align} 
In other words, $Y_j$ is a doubly stochastic random variable (a normal distribution whose mean and variance are themselves random variables).

\section{Application to Precoder Scheme} \label{sec:model}
\begin{figure*}
\centering 
\includegraphics[width=1.1\textwidth]{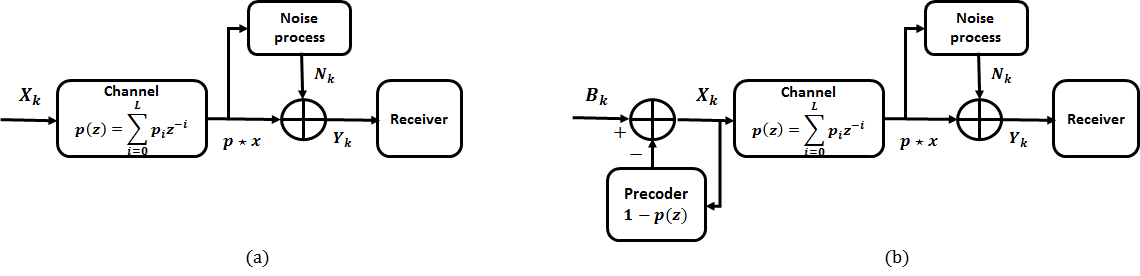}
\vspace{-1.1cm}
\caption{Block diagram of the communication system (a): without precoder, (b): including precoder.}
\label{fig.block} %% label for entire figure
\vspace{-0.5cm}
\end{figure*}

Having allowed for negative input signal while keeping the system linear, our goal is to design a proper precoder to eliminate ISI for the channel described in equation \eqref{noprecoder-1} and \eqref{noprecoder}. Equation \eqref{noprecoder-1} can be written as $Y=p\star X + N$, where $X$ denotes the sequence $\{X_k\}$ and $Y$ denotes the sequence $\{Y_k\}$. Taking the $z$-transform from both sides, we obtain $\hat{Y}(z)=\hat{p}(z)\hat{X}(z)+\hat{N}(z)$, where $\hat{p}(z) = \sum_{i=0}^{L}p_i z^{-i}$. See Fig. \ref{fig.block} (a). To design a precoder for this system, we  add a precoder block at the transmission side, as depicted in Fig. \ref{fig.block} (b). In this scenario the input sequence $B=\{B_k\}$ is first fed into a causal realization of the inverse channel (precoder), i.e., $1/\hat{p}(z)$, to produce the transmission symbols of $X_k$. Because $1/\hat{p}(z)$ is the inverse of the system $\hat{p}(z)$, $B=p\star X$, \emph{i.e.,} 
\begin{align}\sum_{k=0}^{L} p_k X_{j-k}=B_j.\label{eq:B_j}\end{align}
This is done to cancel the ISI at the receiver side. Using equation \eqref{eq:B_j}, we can rewrite \eqref{noprecoder} as follows: 
$Y_j =B_j+N_j$ where $N_j \sim \mathcal{N}(0, \frac{B_j}{V_R})$. In other words,
\begin{align}\label{precoder}
Y_j \sim \mathcal{N}(B_j, \frac{B_j}{V_R}),
\end{align}
with the ISI completely canceled on the receiver side. 
Of course, ISI cancellation is achieved assuming perfect reaction, \emph{e.g.,} $\msKappa=\infty$ in \eqref{reaction}. However, as we argue in Section \ref{section:imperfect-reaction}, an imperfect reaction comes close to a perfect reaction (in a sense that will be clarified later) exponentially fast in $\msKappa T_r$ where $T_r$ is the duration of the reaction. 

 %%%%%%%%%%%%%
%%%%%%%%%%%%%
%
%

\subsection{Power consumption and stability of the precoder}\label{sec:stability}
Assume a simple BPSK modulation with i.i.d.\ and uniform $\{ B_k \} \in \{ -\beta, \beta \}$. Then, the transmitted symbols $X_k$'s are produced by passing $B_k$'s through the precoder. We need to compute $\E \{ |X_k|\}$ for the average consumed transmission power. Furthermore, it is important to investigate the stability and causality of the precoder filter. 

Let us begin with showing the stability of the precoder. We need to prove that all of the poles and zeros of $1/\hat{p}(z)$ are inside the unit circle. Observe that
 \begin{align}
 \frac{1}{\hat{p}(z)} = \frac{1}{\sum_{i=0}^{L}p_i z^{-i}}=\frac{z^L}{\sum_{i=0}^{L}c_iz^{i}}
 \end{align}
 where $c_i= p_{L-i}$. We have $c_0<c_1<\cdots<c_L$ by equation \eqref{condition}. Therefore it is enough to show that all of the roots of $\sum_{i=0}^{L}c_iz^{i}$ lie in the unit disk. 
 %%%
 %%%
 %%%
To show this, we use the following theorem:
\begin{theorem} (Enestrom-Kakeya\cite{bound}) \label{Th:1}
If all coefficients of the polynomial
$g(x) = c_{n-1} x^{n-1} + \cdots + c_{0}$ are positive reals, then for any (complex) root $\zeta$ of this
polynomial, we have
\begin{align}
\min_{1\leq i \leq n-1} \{  \frac{c_{i-1}}{c_i} \}  \leq |\zeta | \leq \max_{1\leq i \leq n-1} \{  \frac{c_{i-1}}{c_i} \}.
\end{align}
\end{theorem}

Considering Theorem \ref{Th:1}, since $\max_{i} \{  \frac{c_{i-1}}{c_i} \}<1$, all roots of $\sum_{i=0}^{L}c_iz^{i}$ have absolute value of smaller than $1$.
Therefore all poles of $1/\hat{p}(z)$ lie inside unit disk, \emph{i.e.}, the filter is bounded input/bounded output stable. Thus, stability provides a new reason for choosing of the sampling time $T_s$ as in equation \eqref{eqn:defTs}.

Now, let us turn to the average power consumption of the precoder. While finding an analytical expression for $\E \{ |X_k|\}$ seems difficult, one can find an upper bound for $\E \{ |X_k|\}$. Observe that
$\E \{ |X_k|\}\leq \sqrt{\E \{X_k^2\}}$. Thus, we need an upper bound on the energy of $X_k$. The i.i.d.\ sequence of $B_k$ is a stationary discrete random process, which is passed through the filter $1/\hat{p}(z)$. Therefore, the autocorrelation function and $\E \{X_k^2\}$ is finite and can be computed. 
%%%%%%%%%%%%%%% 
%%%%%%%%%%%%%%%

\subsection{Imperfections due to  Diffusion Coefficient Mismatch and Quantization}\label{quan}

\textbf{Diffusion coefficient mismatch:} 
This occurs when $D_A\neq D_B$. As mentioned in Example \ref{example1n} and shown in Section \ref{Numeric} via numerical simulations, coefficient mismatch of up to 10 percent does not significantly degrade the error performance. While no explicit solution is known in general, for the case of fast reaction $\msKappa\rightarrow\infty$, the differential equations \eqref{eq1av} and \eqref{eq2av} reduce to the ones for instantaneous reaction \cite{Fast1}, with 
$$\rho_A(t,\vec{r})=\max(\rho(t,\vec{r}), 0), $$
$$\rho_B(t,\vec{r})=\max(-\rho(t,\vec{r}), 0), $$
with $\rho(t,\vec{r})$ satisfying 
\begin{align}
\frac{\partial \rho(t,\vec{r})}{\partial t} &= \nabla ^2 \Phi(\rho(t,\vec{r})) -  \nabla\cdot  (v(\vec{r}) \rho(t,\vec{r}))+s(t,\vec{r}),
\label{eq01nn} 
\end{align}
where $\Phi(x)=D_Ax$ for $x\geq 0$ and $D_Bx$ for $x\leq 0$. When $v(\vec{r}) \rho(t,\vec{r})=0$, this is a special case of the generalized porous medium equation \cite[Eq. (2.10)]{Vazquez}  (see also \cite{Fast1} and the references cited for existence, uniqueness, stability of this equation). Approximate solutions for this problem has been found \cite{Fast4} for certain boundary conditions. Interestingly, like our simulation result given in  Section \ref{Numeric}, it can be observed from \cite[Fig 1.]{Fast4} that a mismatch of up to 10 percent does not significantly vary the movement of reaction interface. 

\textbf{Quantization:} Even if the input of the precoder is binary, the output of the precoder will have multiple levels; this may make the transmission scheme complicated. As a result, we can consider quantization of input $X_k$ before its transmission. This can reduce the complexity of the transmitter. 

Generally, we can define a quantizer consisting of a set of intervals $\mathcal{F}= \{ f_i; i=1,2,3,\cdots, M\}$ and a set of levels $\mathcal{L} =\{l_i, i=1,2,3,\cdots, M\}$, so that the quantizer $\mathcal{Q}$ (often called the quantization rule) can be expressed as
\begin{align}
\mathcal{Q}(x) = \sum_{i=1}^M c_i \mathbb{1} _{f_i}(x),
\end{align} 
where the indicator function is given by
\begin{align} \label{indicator}
 \mathbb{1} _{f_i}(x)= \left\{ 
  \begin{array}{l l}
    1  &: \text{if}~x \in f_i ,  \\
    0  &:  \text{if}~x \notin f_i.  \\
  \end{array} \right.
\end{align}
The transmitter sends $[{X}_k]_{\mathcal{Q}}=\mathcal{Q}(X_k)$ instead of $X_k$. Thus, $e_k=X_k-[{X}_k]_{\mathcal{Q}}$
is the quantization error. Then, the concentration of molecules around the receiver equals $$p\star [{X}]_{\mathcal{Q}}=Y-p\star e.$$
The second term $p\star e$ is the quantization error at the receiver. Note that $$p\star e=\sum_j p_je_{i-j}\leq (\sum_j p_j)e_{\max}$$ where $e_{\max}$ is the maximum quantization error. Thus, the output error term $p\star e$ vanishes as we increase the  quantization levels and reduce $e_{\max}$. However, given a fixed number of quantization levels, we can ask for the best possible quantization intervals and levels. While the simplest quantizer is the \emph{uniform} quantizer, non-uniform quantizers have better performance in general. The literature on quantization generally measures the quality of a quantizer by the distance between the input and output of the quantizer. An example of a distance function is the mean squared error (MSE):
\begin{align}\label{distortion}
\text{D}\left(\mathcal{Q}\right) = \E\left[ (X-\mathcal{Q}(X))^2\right],
\end{align}
where $\text{D}(\mathcal{Q})$ is the average distortion, $\E [\cdot]$ denotes the expectation operator over the density function of the random variable $X$ given to the quantizer. The optimum quantizer is then found to minimize the average distortion (\ref{distortion}) over all valid quantization rules, \emph{i.e.},
\begin{align} \label{opt_quan}
\mathcal{Q}^{*} = \argmin_{Q} \text{D}\left(Q\right).
\end{align}
For the  mean squared distance defined in (\ref{distortion}), the optimum quantizer of (\ref{opt_quan}) is called the Lloyd quantizer \cite{Lloyd}, and is non-uniform in generals. 

For the scenario considered in this paper, minimizing (\ref{distortion}) does not generally lead to minimize symbol error rate (SER) for the transceiver link. That is, although Lloyd quantizer has the maximum reduction in distortion between the input and output of the quantization block, it does not have the best performance considering SER metric. This effect is investigated in more details in Section \ref{Numeric} via numerical simulations.

%%%
%%%

%%%%%%%%%%%%%%%%%%%%%%%%%%%%%%

%%%%%%%%%%%%%%%%%%%%%%%%%%%%%%%%%%%%%

%

\section{TS modulation with imperfect reactions}
\label{section:imperfect-reaction}

As mentioned earlier, chemical reaction allows for decreasing the noise variance. Below, we consider slow reaction with large or small reaction times separately. 

\subsection{Compensating slow reaction rate with large reaction time}\label{secslowreact}
Thus far, we considered the case of fast reactions. But one may not always need fast reactions. Weak reaction rates $\msKappa_1$ can be compensated by increasing the reaction time window $T_s$ for it to take effect. Here, we provide a non-rigorous argument to illustrate this point. Let us consider the use of molecule $C$ as in \eqref{reaction} (local reaction at the receiver) and one-way imperfect reaction with some finite $\msKappa<\infty$. For simplicity consider that the environment has no drift. While doing an exact calculation can be difficult, we proceed with an approximate calculation to gain some insight. Molecules are released at the beginning of each time slot of duration $T_s$,  for a period of $T_e<T_s$ seconds. Assume that the receiver samples the environment at the end of time-slot; $T_r<T_s$ seconds prior to that, it releases molecules of type $C$ to clean up and prepare the environment for sampling.  Assume that the inter-transmission interval $T_s$ is large enough for the concentration of molecules of type $A$ and $B$ to  stabilize and reach the steady state values $\bar{\rho}_A$ and $\bar{\rho}_B$ around the receiver, at the time of the release of molecule $C$. Due to this stabilization, the terms $D_A \nabla ^2 \rho_{A}(t,\vec{r})$ and $D_B\nabla ^2 \rho_{B}(t,\vec{r})$ are zero. At this time, the receiver releases molecules of type $C$ to clear up its surroundings. If the concentration of $C$ is assumed to be kept fixed at some level $\bar{\rho}_C$ in its surroundings, equations \eqref{eq1avN} and \eqref{eq2avN} in the vicinity of the receiver simplify to
\begin{align}
\frac{\partial \rho_{A}}{\partial t} &= -\msKappa\rho_A \rho_B\lambda,\label{eq1av2} \\
\frac{\partial \rho_{B}}{\partial t} &= -\msKappa\rho_A \rho_B\lambda,\label{eq2av2}
\end{align}
where $\lambda=\bar{\rho}_C^{\gamma}$. Assuming the initial condition  $\rho_A(0,\vec{r})=\bar{\rho}_A$ and $\rho_B(0,\vec{r})=\bar{\rho}_B$ for any $\vec{r}$ in the vicinity of the receiver, we are interested in computing the concentration of $A$ and $B$ after passage of $T_r$ seconds.

Subtracting \eqref{eq1av2} from \eqref{eq2av2}, we see that $\rho_A(t,\vec{r})-\rho_B(t,\vec{r})$ is invariant in $t$, and thus equal to $\Delta=\bar{\rho}_A-\bar{\rho}_B$. Without loss of generality assume that $\Delta>0$ (the case $\Delta<0$ is similar). Then, equation \eqref{eq1av2}  reduces to
\begin{align}
\frac{\partial \rho_{B}}{\partial t} &= -\msKappa\lambda\rho_B (\Delta+\rho_B)\label{eq1av3}.
\end{align}
Solving this equation, we get
\begin{align}
\rho_B(t,\vec{r})=\frac{\Delta \bar{\rho}_B e^{-\msKappa\lambda\Delta t}}{\Delta+\bar{\rho}_B-\bar{\rho}_B e^{-\msKappa\lambda\Delta t}} \leq \bar{\rho}_B e^{-\msKappa\lambda\Delta t}.
\end{align}
This shows that the concentration of $B$ converges to zero exponentially fast in $t$.  We verify this exponential behavior in Section \ref{Numeric} via numerical simulations.  As a result, as long as
\begin{align}\bar{\rho}_B e^{-\msKappa\lambda\Delta T_r}\ll \Delta\end{align}
we can say that the concentration of molecules of type $B$ is small in comparison with the concentration of molecules of type $A$ at the time of measurement by the receiver. From here, one can conclude the possibility of compensating for small $\msKappa$ by increasing $T_r$; as long as the product $\msKappa T_r$ is large, the error probability of the system will be close to that of perfect reaction $\msKappa=\infty$.

\subsection{Slow reaction rate and reaction time}
We now consider the case that both reaction rate $\msKappa$ and the time-slot duration $T_s$ are small. Let us begin with the extreme case of no reaction amongst molecules. In this case the diffusion system is linear. We can write
$$\rho_A(t, \vec{r}^*)=\sum_{k: X_k>0}X_k\cdot p(t-kT_s).$$
$$\rho_B(t, \vec{r}^*)=\sum_{k: X_k<0}(-X_k)\cdot p(t-kT_s).$$
As a result,
\begin{align*}\rho_A(jT_s, \vec{r}^*)+\rho_B(jT_s, \vec{r}^*)=\sum_{k}|X_k|\cdot p(jT_s-kT_s)
=\sum_{k=0}^j|X_k|\cdot p(jT_s-kT_s).
\end{align*}
Hence, from $p_j = p(jT_s)$, and $p_j=0$ for $j>L$, we obtain
\begin{align}\label{noprecoder2n2}
Y_j \sim \mathcal{N}\left(\sum_{k=0}^{L} p_k X_{j-k}, \frac{1}{V_R}\sum_{k=0}^{L} p_k \left|X_{j-k}\right|\right).
\end{align} 
Comparing this equation with \eqref{noprecoder}, we see that the absolute value is moved inside the summation on the individual terms. 

Now, consider the case of partial reaction. Either of the reactions of the types \eqref{eq1av} and \eqref{eq2av}, or \eqref{reaction}  have a net effect of reducing the concentration of molecules of type $A$ and $B$. However, if this reaction is imperfect, the densities of both of molecules of types $A$ and $B$ can be still positive at the receiver. We need the sum value $\rho_A(jT_s, \vec{r}^*)+\rho_B(jT_s, \vec{r}^*)$ to be able to compute the measurement noise at the receiver. Because diffusion-reaction equations are deterministic, the sum $\rho_A(jT_s, \vec{r}^*)+\rho_B(jT_s, \vec{r}^*)$ is a (non-linear) deterministic function of past transmissions (or equivalently the past $L+1$ transmissions because the channel memory is $L$). We can denote this sum by $g(X_j, X_{j-1}, \cdots, X_{j-L})$. Then, we will have
\begin{align}\label{noprecoder3n3}
Y_j \sim \mathcal{N}\left(\sum_{k=0}^{L} p_k X_{j-k}, \frac{1}{V_R}g(X_j, X_{j-1}, \cdots, X_{j-L})\right).
\end{align} 
With this notation, for ``full reaction" and ``no reaction", we have
$$g_{\mathsf{full-reaction}}(x_j, x_{j-1}, \cdots, x_{j-L})=\left|\sum_{k=0}^{L} p_k x_{j-k}\right|$$
$$g_{\mathsf{no-reaction}}(x_j, x_{j-1}, \cdots, x_{j-L})=\sum_{k=0}^{L} p_k \left|x_{j-k}\right|.$$
To find $g(X_j, X_{j-1}, \cdots, X_{j-L})$ one has to solve the non-linear differential reaction-diffusion equations to find  the individual concentrations $\rho_A(jT_s, \vec{r}^*)$ and $\rho_B(jT_s, \vec{r}^*)$. For instance, consider equations \eqref{eq1av} and \eqref{eq2av}.  The difference $\rho=\rho_A-\rho_B$ follows a linear differential equation and can be solved for. Using the fact that $\rho_B=\rho_A-\rho$, we get the following differential equation for $\rho_A$,
\begin{align}
\frac{\partial \rho_{A}}{\partial t} &= D_A \nabla ^2 \rho_{A} -  \nabla\cdot  (v \rho_A)
-\msKappa\rho_A (\rho_A-\rho)+s_{A}.\label{eq1avnew1}
\end{align}
There is a rich literature on reaction-diffusion equations. While no explicit analytical solution is known for them, the literature provides thorough studies of specific reaction-diffusion equations; these special cases provide insights into the complexities of such differential equations.\footnote{One of these special cases, the Fisher's equation, can be understood as a special case of \eqref{eq1avnew1}: assume $s_A(t, \vec{r})=v(t, \vec{r})=0$, and the difference of the concentrations is equal to one, $\rho(t, \vec{r})=1$ for all $t, \vec{r}$. Then, \eqref{eq1avnew1} reduces to 
\begin{align}
\frac{\partial \rho_{A}}{\partial t} &= D_A \nabla ^2 \rho_{A} +\msKappa\rho_A (1-\rho_A).\end{align}
This is the Fisher's equation and is known to have a \emph{traveling wave solution}. This highlights how different can $\rho_A$ and  $\rho$ behave; here the difference of concentrations, $\rho$, is a constant function while $\rho_A$ is a wave function.}

Despite lack of explicit analytical solution for the non-linear diffusion-reaction equation, it is possible to prove bounds for $g(x_j, x_{j-1}, \cdots, x_{j-L})$. It is intuitive that the noise variance should be between that of no-reaction and full reaction, \emph{i.e.,}
$$\left|\sum_{k=0}^{L} p_k x_{j-k}\right|\leq g(x_j, x_{j-1}, \cdots, x_{j-L})\leq \sum_{k=0}^{L} p_k \left|x_{j-k}\right|.$$
The inequality $$\left|\sum_{k=0}^{L} p_k x_{j-k}\right|\leq g(x_j, x_{j-1}, \cdots, x_{j-L})$$
easily follows from the fact that the sum $\rho_A(jT_s, \vec{r}^*)+\rho_B(jT_s, \vec{r}^*)$ is always greater than or equal to $|\rho_A(jT_s, \vec{r}^*)-\rho_B(jT_s, \vec{r}^*)|$. To prove the inequality 
\begin{align}g(x_j, x_{j-1}, \cdots, x_{j-L})\leq \sum_{k=0}^{L} p_k \left|x_{j-k}\right|,\label{eqn:inequ-left-g}\end{align}
we prove the following result:
\begin{theorem}\label{thm2d}In an environment with no flow velocity $v=0$, the function $g(x_j, x_{j-1}, \cdots, x_{j-L})$ is subadditive, \emph{i.e.,} 
\begin{align}
g(x_j+x'_j, x_{j-1}+x'_{j-1}, \cdots, x_{j-L}+x'_{j-L}) 
\leq g(x_j, x_{j-1}, \cdots, x_{j-L})+g(x'_j, x'_{j-1}, \cdots, x'_{j-L}),
\end{align}
holds for any arbitrary $(x_j, x_{j-1}, \cdots, x_{j-L})$ and $(x'_j, x'_{j-1}, \cdots, x'_{j-L})$.
\end{theorem}
The proof is given in Appendix A. 

One can deduce \eqref{eqn:inequ-left-g}  from this theorem as follows: take an arbitrary tuple $(x_j, x_{j-1}, \cdots, x_{j-L})$. Let $y_i=x_i \mathbf{1}[x_i\geq 0]$ and $z_i=x_i \mathbf{1}[x_i\leq 0]$ for $j-L\leq i\leq j$, where $\mathbf{1}[\cdot]$ is the indicator function. Then,
\begin{align}
g(x_j, x_{j-1}, \cdots, x_{j-L})
&\leq g(y_j, y_{j-1}, \cdots, y_{j-L})+g(z_j, z_{j-1}, \cdots, z_{j-L})\nonumber
\\&=g_{\mathsf{no-reaction}}(y_j, y_{j-1}, \cdots, y_{j-L})+g_{\mathsf{no-reaction}}(z_j, z_{j-1}, \cdots, z_{j-L})\label{eqn:kjfnb1}
\\&=\sum_{k=0}^{L} p_k \left|y_{j-k}\right|+\sum_{k=0}^{L} p_k \left|z_{j-k}\right|\nonumber
\\&=\sum_{k=0}^{L} p_k \left|x_{j-k}\right|,\label{eqn:kjfnb234}
\end{align}
where \eqref{eqn:kjfnb1} follows from the fact that in $g(y_j, y_{j-1}, \cdots, y_{j-L})$ all $y_i$'s are positive and hence only molecules of type $A$ are released (hence no reaction may occur). Similarly, in $g(z_j, z_{j-1}, \cdots, z_{j-L})$ only molecules of type $B$ are released, hence there is no possibility of any reactions. Finally, \eqref{eqn:kjfnb234} follows from $|y_i|+|z_i|=|x_i|$ for $j-L\leq i\leq j$.

%****
%****
%****
%$$$$$$$$$$$$$$$$$$$$$$$$$$$$$$$$$$$$$$$$$$$$$
%%%%%%%%%%%%%%%%%%%%%%%%%%%
%*****************************************************************************
\section{Numerical Results} \label{Numeric}
For numerical results we consider equiprobable binary sequences and three modulations and signaling of CSK, MCSK, and the proposed TS signaling with precoder. In CSK for macroscale molecular communication (amplitude modulation), only one molecule type is used and  information bits are encoded in the concentration of the released molecules into the medium. In order to transmit the bit $1$, the transmitter releases a fixed (say $\beta$) molar of molecules of a give type. To transmit the bit $0$, no molecules are released. In the MCSK scheme, similar to CSK, same amount of molecules are released into the medium for each information bit, with the difference that two types of molecules, \emph{i.e.}, $A$ and $B$ are utilized for odd and even time slots, respectively. 

One can consider the CSK and MCSK schemes for two extreme cases: one for \emph{no-memory} and the other for \emph{infinite-memory} at the receiver \cite{me1}. For the no-memory scheme, receiver has no memory unit and hence cannot remember the previous transmissions, \emph{i.e.}, no knowledge about the current ISI. In the infinite-memory scheme, we suppose that at the end of each time slot, a \emph{genie} would inform the receiver about the exact value of the current ISI due to the previous transmissions.\footnote{This is equivalent to assume that all prior transmissions are correctly decoded at the receiver and known by the receiver.} While exact analytical expressions are not previously reported for a genie-aided CSK and MCSK, we used the MAP rule and derived the optimum detector for each transmission, and used the explicit expressions in our plots.

 We also consider transmission in a one-dimensional unbounded environment with zero boundary conditions at plus and minus infinity. As we saw earlier, transmissions waveforms should be H\"older continuous. For consistency with the existing literature on molecular communication, we assume instantenous release of molecules (delta function); as commonly mentioned in engineering textbooks, one can find a sequence of H\"older continuous functions that approximate the delta function.
For fair comparisons between the proposed scheme and the other ones, we keep the expected total sum of the released concentration of molecules of type $A$ and $B$  in all schemes equal to each other and equal to $\beta$.
The system parameters are $D= 2.2\times 10^{-9} m^2/s, V_R = 5\times10^{-16}cm^3$, and $\|r^{*}\| = 2.15\times 10^{-7}m$. 

\begin{figure} % Fig. 4
\centering 
\includegraphics[width=0.65\textwidth]{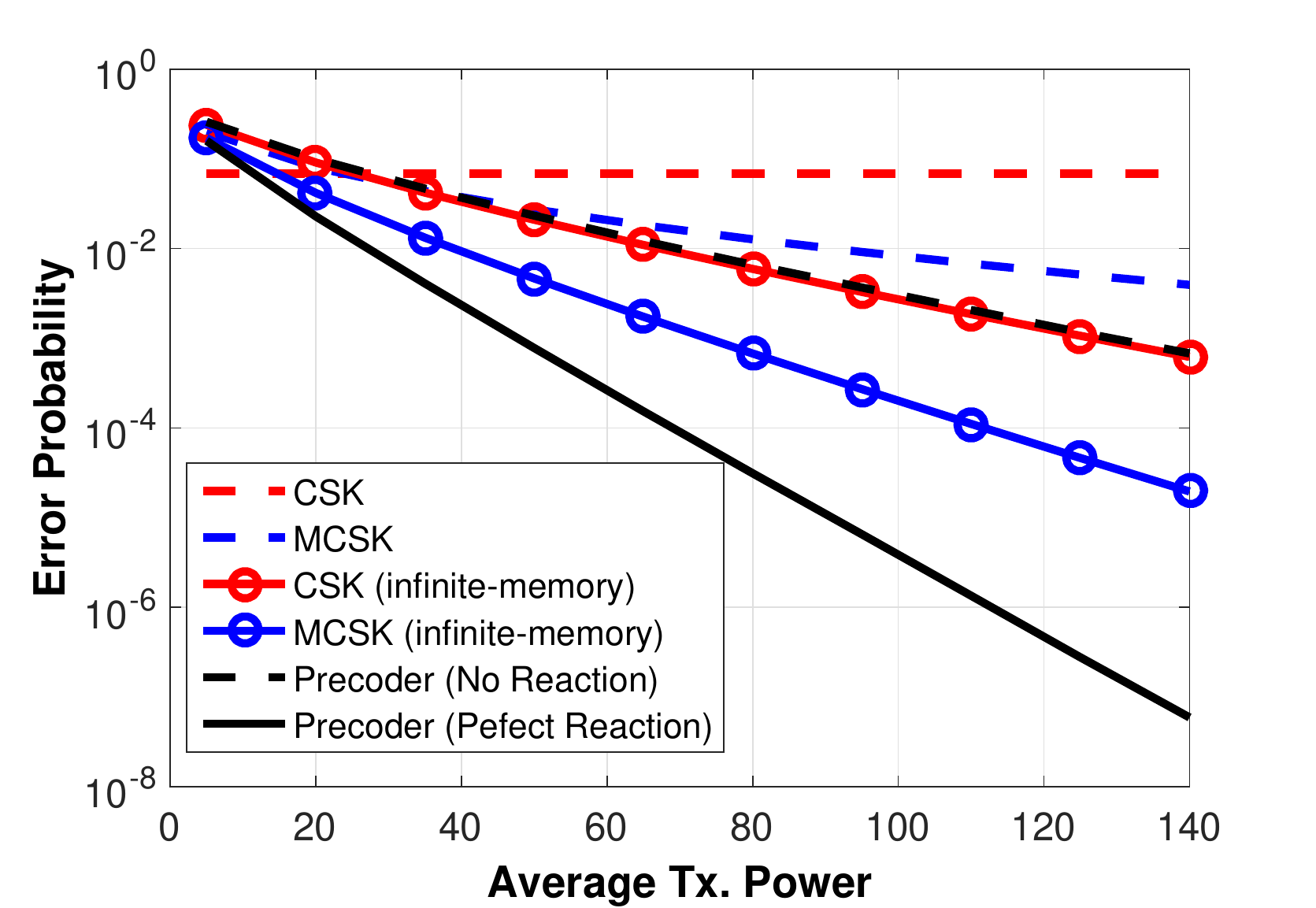}
\vspace{-0.3cm}
\caption{The performance comparison among proposed precoder scheme, CSK, and MCSK modulations for different values of average transmission power. }
\label{fig.comp} 
\vspace{-0.5cm}
\end{figure}
%****************************************************************************
\begin{figure} % Fig. 5
\centering
\subfigure[]{
 \centering
\includegraphics[width=0.63\linewidth]{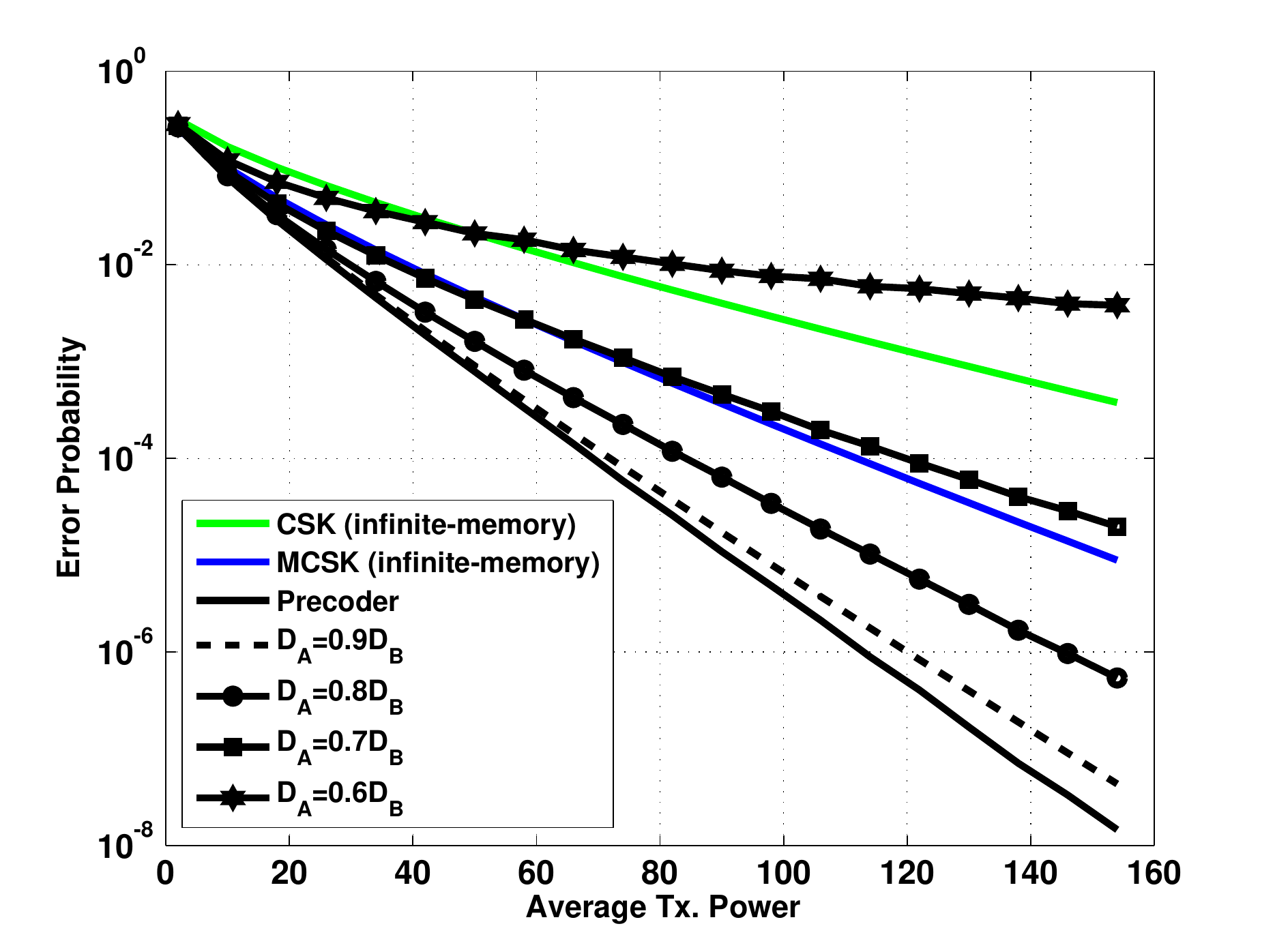}    
}
\vspace{-1.0cm}
\subfigure[]{
\centering
\includegraphics[width=0.63\linewidth]{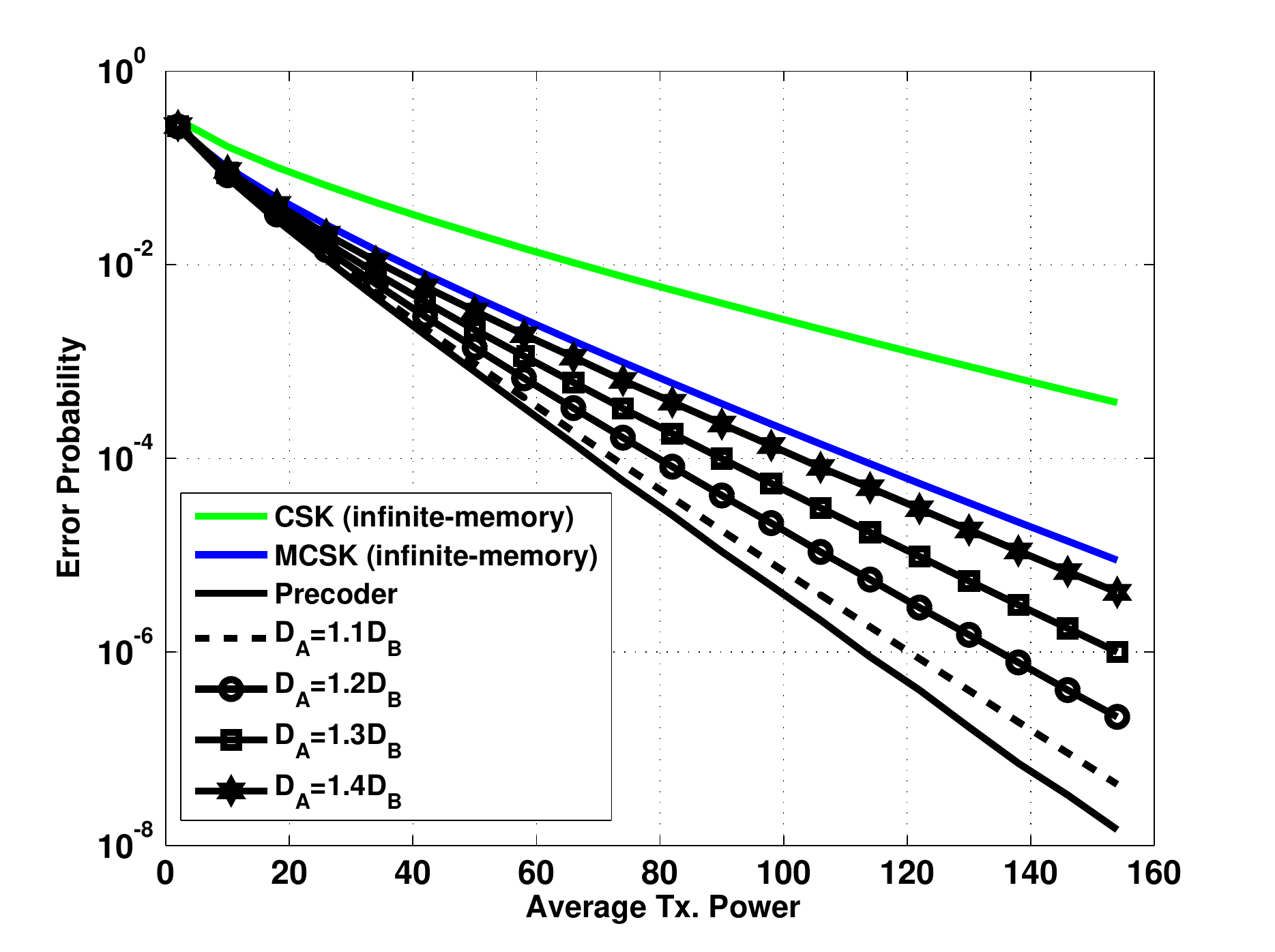}      
}
\vspace{0.8cm}
\caption{\footnotesize{The performance comparison among proposed precoder scheme, CSK and MCSK modulations for different values of average transmission power and different diffusion coefficients for (a): $D_B<D_A$, and (b): $D_B>D_A$. }
}
\label{fig.DiffA}
\vspace{-0.5cm}
\end{figure}
In Fig. \ref{fig.comp}, the probability of error for all modulations are compared with each other versus the concentration of transmitted molecules (we denote an scale of this value by power). As Fig. \ref{fig.comp} depicts, higher error is for the CSK modulation which is a simple modulation scheme that uses one molecule type to transmit information bits. In addition, it can be observed that the proposed TS scheme (with precoder and assuming $\msKappa=\infty$) significantly outperforms other modulation and signaling schemes, even with infinite memory at the receiver. On the other hand, our proposed precoder has a more complicated design compared to the CSK or MCSK schemes. Furthermore, in CSK and MCSK, we have only two transmission levels, but in the one with precoder, transmitter is assumed to be able to diffuse at any arbitrary level (subject to the average transmission power). 

Fig. \ref{fig.DiffA} depicts the performance when we have a mismatch in diffusion coefficients. As expected, once the value of $D_B/D_A$ deviates from $1$, the probability of error increases. However, this increase in error probability is relatively negligible if the mismatch between $D_A$ and $D_B$ is less than 10 percent.

\begin{figure}
\centering
\subfigure[]{
 \centering
\includegraphics[width=0.60\linewidth]{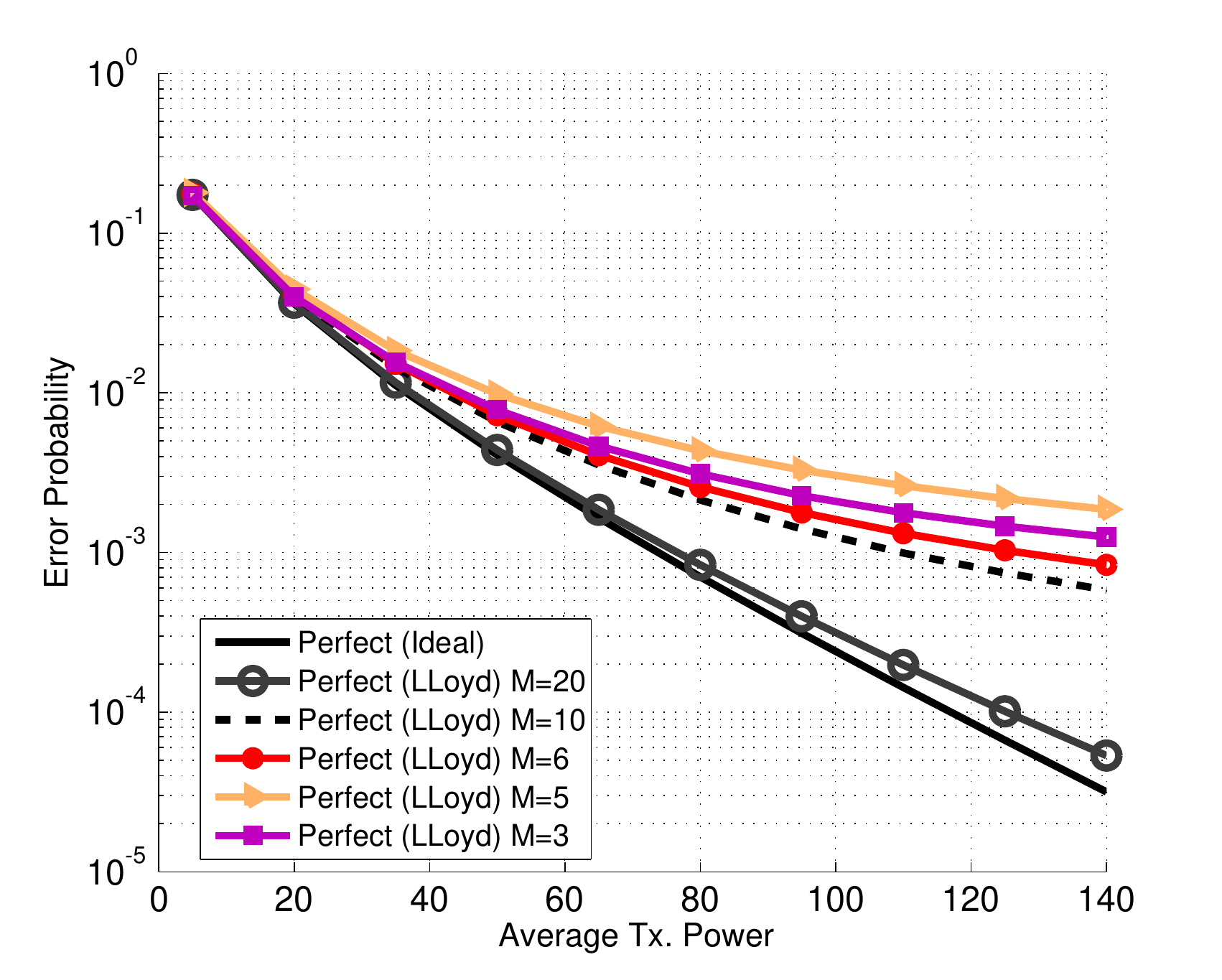}      
}
\vspace{-0.2cm}

\subfigure[]{
\centering
\includegraphics[width=0.60\linewidth]{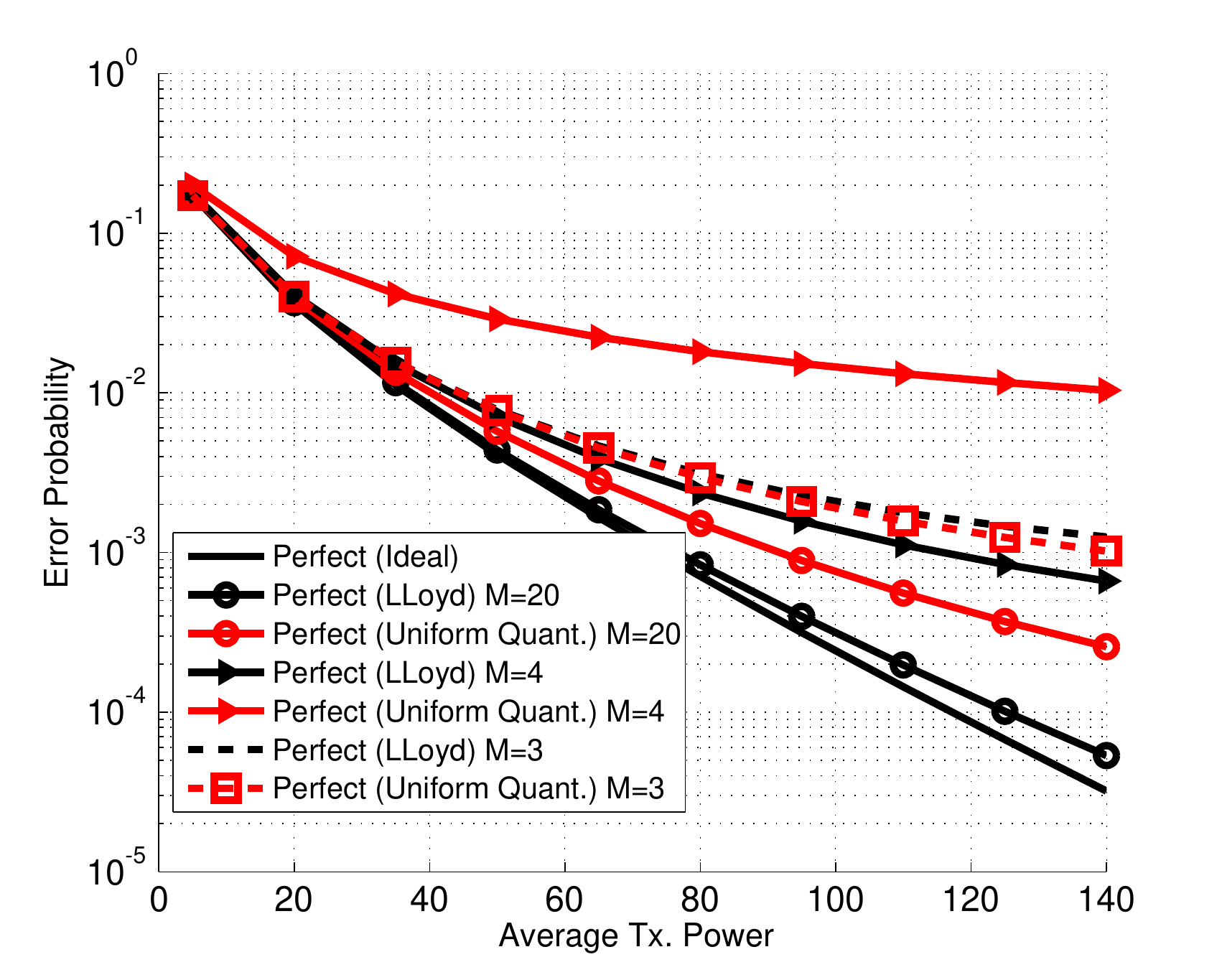}      
}
\vspace{-0.2cm}
\caption{\footnotesize{The effect of quantization on the precoder scheme with complete reaction ($\msKappa=\infty$) utilizing (a): Lloyd algorithm, and (b): Lloyd and uniform algorithms. }
}
\label{fig.quantize}
\vspace{-0.5cm}
\end{figure}

To make design of the precoder simpler, we investigate the impact of quantizer in Fig. \ref{fig.quantize} (a) versus the average number of released molecules (power), utilizing Lloyd algorithm. It is clear that as the number of quantization levels ($M$) increase, the MMSE of quantizer decreases. However, this does not necessarily improve the performance of the transceiver link.
That is, from Fig.  \ref{fig.quantize} it can be realized that although by increasing the quantization levels the average distortion decreases, the error rate does not decrease. For instance, the error probability for the quantization level of $M=3$ is less than the one for $M=5$. This phenomenon is not strange as Lloyd quantizer just minimizes the MMSE and not the end-to-end error probability. For a better comparison, in Fig. \ref{fig.quantize} (b) we have plotted the error probability for different quantization levels for Lloyd and uniform qunatization rules. Although for large quantization levels Lloyd quantizer outperforms uniform quantizer, for some small levels, uniform quantizer can outperform the Lloyd quantizer.

To evaluate the effect of incomplete reaction on the performance of receiver, we investigate the affection of forward reaction rate on the number of received molecules $A$ and $B$ at the receiver. To this end, we consider the system of equations given in (\ref{eq1av})-(\ref{eq2av}) for  a one-dimensional medium. To solve the system of equations, we employ the finite difference method (FDM) \cite{FDM1}, \cite{Gold}.
 In this method, the x-axis and time are discretized into infinitesimal intervals of $\Delta x$ and $\Delta t$, respectively, and a set of recursive equations are derived accordingly. 

\begin{figure}
\centering 
\includegraphics[width=0.65\textwidth]{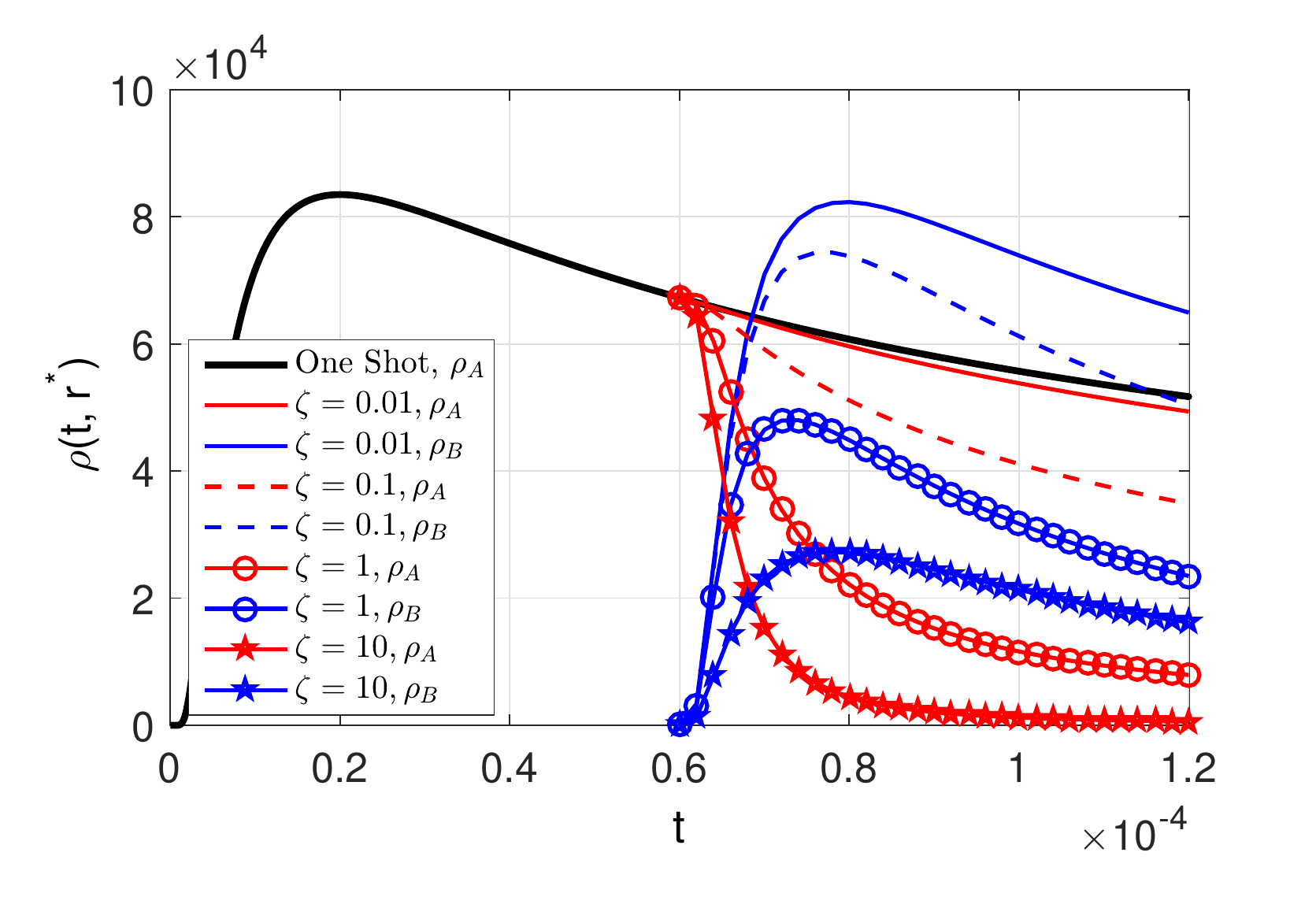}
\vspace{-0.9cm}
\caption{The impact of kinetics of reaction on a realization of transmission scheme for releasing $3\times 10^{6}$ molecules of type $A$ from origin at time zero, and then $3\times 10^{6}$ molecules of type $B$ at time  $6\times 10^{-5}$. Here, $T_s = 6T_r = 6 \times 10^{-5} sec$. }
\label{fig.realization} %% label for entire figure
\vspace{-0.4cm}
\end{figure}

Fig. \ref{fig.realization} depicts the concentration of molecules $A$ and $B$ for two consecutive transmission of $A$ and then $B$, for different values of the reaction rate. In particular $\beta=3\times 10^{6}$ molecules of type $A$ are released at time zero, and then $\beta=3\times 10^{6}$ molecules of type $B$ are released at time $T_s = 6\times 10^{-5} sec$. 
The receiver volume is equal to $0.01 \mu m$ and the distance between the transmitter and receiver is $0.21 \mu m$. As the reaction rate increases, although the difference between the concentration of molecule $A$ and $B$ is constant, more molecules make reaction and disappear from the medium. Focusing on the second transmission, it can be seen that for the reaction with $\msKappa > 10$, the limiting reactant (molecule $A$) is almost completely removed from the medium at $t= 2T_s = 12 \times 10^{-5}$. This verifies the calculations presented in Section \ref{secslowreact} that for reaction rate larger than a threshold, the concentration of the smaller reactant degrades exponentially fast.
%****************************************************************************
%%%%%%%%%%%%%%%%%%%%%%%%%%%%%%%%%%
\begin{figure}
\centering 
\includegraphics[width=0.65\textwidth]{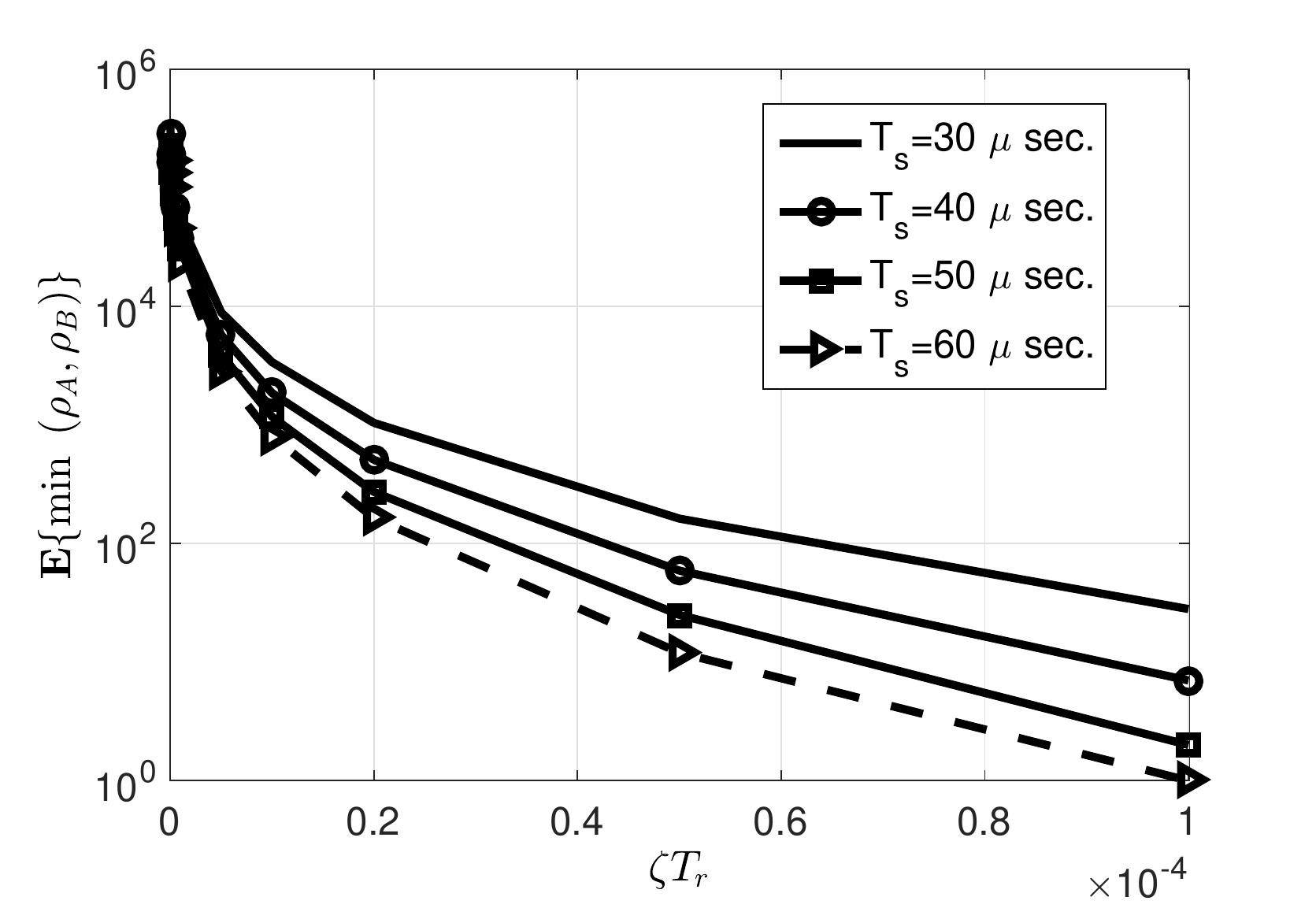}
\vspace{-0.4cm}
\caption{The impact of kinetics of reaction on the mean number of received molecules at the receiver for the input values to the precoder with amplitudes of $3\times 10^{6}$.}
\label{fig.imperfect2} %% label for entire figure
\vspace{-0.4cm}
\end{figure}
%******************************
%

Fig. \ref{fig.imperfect2} verifies the intuitive argument given in Section \ref{secslowreact}. It shows the effect of  reaction rate, $\msKappa$, on the mean concentration of received limiting molecules at the receiver, i.e., $\E \{ \min ( \rho_A(jT_s,r^{*}), \rho_B(jT_s,r^{*}) ) \}$ where the expectation is made over 1000 trials. 
Since we have used precoder, solving (\ref{diff}) gives the difference of the concentrations at the receiver at the end of each time slot equal to $9\times 10^{4}$. On the other hand, since the reaction is not complete, the concentration of limiting reactant at the receiver is not zero. It can be observed that as the reaction rate multiplied by the duration of reaction, i.e., $\msKappa T_r$ linearly increases, the number of received limiting molecules decreases exponentially fast.

%%*****
\section{Conclusion}\label{sec:conc}
The restriction of positive signals makes handling of ISI a challenge for molecular communication. In this paper, we proposed a new modulation scheme, named TS, to simulate negative signals. The modulation scheme places information on the {difference} of concentrations of $A$ and $B$ regardless of whether a chemical reaction is present or not. The difference of concentrations is a ``reaction-rate invariant" of chemical reactions such as 
$A+B\xrightarrow{\msKappa} Outputs$ or
$A+B+\gamma C\xrightarrow{\msKappa} Outputs$, meaning that the difference of concentrations follows a differential equations that is not affected by the reaction rate $\msKappa$. The key  property used here is that the stoichiometric coefficients of $A$ and $B$ are equal to each other in the above equations. On the one hand, this implies that the difference of concentrations follows a linear differential equation. On the other hand, invariance with respect to reaction rate implies that  the modulation scheme is \emph{robust} with respect to factors that influence the reaction rate such as existence of a  catalyst. However, all of this comes at the cost of losing a degree of freedom by placing information on the difference of concentrations, and not utilizing the individual concentrations. The general idea of using the reaction rate invariants as signalling variables may be extended to other chemical reactions and employed in other communication settings. 

We showed that TS modulation allows for negative inputs, and hence we can employ classical ideas based on channel invertion to introduce a precoder at the transmitter side.  We also discussed the effect of imperfections such as quantization and mismatch between diffusion coefficients of reactants on the performance of the proposed precoder. Numerical results indicate that the TS scheme can strictly outperform previously proposed schemes, even with slow reaction rate and other mentioned imperfections.

%%
%% ********Appendix
\appendices
%\appendix

\section{Some comments on the reaction diffusion equations given in \eqref{eq1av} and \eqref{eq2av}} 
\label{somecomment}
Consider the reaction diffusion equations given in \eqref{eq1av} and \eqref{eq2av}. 
When $\msKappa=0$, \emph{i.e.,} when the non-linear term does not exist, the system whose inputs are $s_A$ and $s_B$ and whose outputs are $\rho_A$ and $\rho_B$ is linear. As a result multiplying both $s_A$ and $s_B$ by a constant $\theta$ would affect $\rho_A$ and $\rho_B$ by the multiplicative constant $\theta$. To understand the case of $\msKappa>0$, it is helpful to consider the following change of variables: $\rho_A=\theta \rho'_A$, $\rho_B=\theta \rho'_B$, $v'=\varphi v$, $D'_A=\varphi D_A$, $D'_B=\varphi D_B$, $s'_A=\theta s_A$, $s'_B=\theta s_B$, and $t=\varphi t'$ in equations \eqref{eq1av} and \eqref{eq2av} for some positive $\theta$ and $\varphi$. Then, after the rescaling, equations \eqref{eq1av} and \eqref{eq2av} become
\begin{align}
\frac{\partial \rho'_{A}}{\partial t'} &=  D'_A \nabla ^2 \rho'_{A} -  \nabla\cdot  (v' \rho'_A)
-\varphi\theta\msKappa\rho'_A \rho'_B+\varphi s'_{A},\label{eq1avred} \\
\frac{\partial \rho'_{B}}{\partial t'} &= D'_B \nabla ^2 \rho'_{B} -  \theta \nabla\cdot  (v' \rho'_B)
-\varphi\theta\msKappa\rho'_A \rho'_B+\varphi s'_{B}. \label{eq2avred}
\end{align}
Note that this is the same reaction-diffusion equation with $\msKappa'=\varphi\theta\msKappa$. Let us begin with the case of $\varphi=1$, $\theta>0$. We observe that multiplying both $s_A$ and $s_B$ by a constant $\theta$ would affect $\rho_A$ and $\rho_B$ by the same multiplicative constant $\theta$ \emph{only if $\msKappa$ is also multiplied by $\theta$.} When we employ $\theta>1$ times higher molecule production rates $s_A$ and $s_B$, we should make the chemical reaction $\theta$ times faster if we want the concentrations $\rho_A$ and $\rho_B$ to be multiplied by $\theta$. If we keep using the same reaction rate $\msKappa$, molecules of type $A$ and $B$ would not cancel out as fast as $\theta\msKappa$ and concentrations of molecules of type $A$ and $B$ will be multiplied by bigger multiplicative constants than $\theta$. This is an effect of the non-linear term.

Finally, consider the case of $\theta=1$, $\varphi>0$. Observe the particular change of variable $t=\varphi t'$ used. Here, we see the intuitive relation between $\msKappa$ and time: if we multiply $\msKappa$, $D_A$ and $D_B$ by two, we are making the reaction and diffusion twice as fast and we expect the same diffusion pattern in half the time. Now, consider the use a time-slotted communication strategy with time steps of size $T_s$. The above discussion shows that decreasing $T_s$ by a factor of two while at the same time increasing $\msKappa$, $D_A$ and $D_B$ by a factor of two would keep the molecular concentrations (and the error probabilities) the same. 

\section{Proof of Theorem \ref{thm2d}} Consider the differential equations
\begin{align}
\frac{\partial \rho_{A}}{\partial t} &= D_A \nabla ^2 \rho_{A} 
-\msKappa\rho_A \rho_B+s_{A},\label{eq1avv} \\
\frac{\partial \rho_{B}}{\partial t} &= D_B \nabla ^2 \rho_{B} 
-\msKappa\rho_A \rho_B+s_{B}. \label{eq2avv}
\end{align}
Let $(\rho_A, \rho_B, s_A, s_B)$ and $(\rho'_A, \rho'_B, s'_A, s'_B)$ be two solutions to the above equations, where $s_A, s'_A, s_B$ and $s'_B$ are assumed to be H\"older continuous functions. Then by adding up equation \eqref{eq1avv} for these two solutions, we obtain 
\begin{align}
\frac{\partial (\rho_{A}+\rho'_A)}{\partial t} &= D_A \nabla ^2 (\rho_{A}+\rho'_A) 
-\msKappa\rho_A \rho_B-\msKappa\rho'_A \rho'_B+s_{A}+s'_A\nonumber \\
&\geq D_A \nabla ^2 (\rho_{A}+\rho'_A) 
-\msKappa(\rho_A+\rho'_A)(\rho_B+\rho'_B)+s_{A}+s'_A,\label{eq1avev2} 
\end{align}
where in \eqref{eq1avev2}, we used the fact that $\rho_A, \rho'_A, \rho_B$ and $\rho'_B$ are non-negative. Similarly,
\begin{align}
\frac{\partial (\rho_{B}+\rho'_B)}{\partial t} &\geq D_B \nabla ^2 (\rho_{B}+\rho'_B) 
-\msKappa(\rho_A+\rho'_A)(\rho_B+\rho'_B)
+s_{B}+s'_B.\label{eq1avev232} 
\end{align}
Furthermore, $(\rho_A+\rho'_A, \rho_B+\rho'_B)$ have the zero initial and boundary conditions. 
Now, consider the reaction-diffusion set of equations for molecule production rates $s_A+s'_A$ and $s_B+s'_B$:
\begin{align}
\frac{\partial \tilde\rho_{A}}{\partial t} &= D_A \nabla ^2 \tilde\rho_{A} 
-\msKappa\tilde\rho_A \tilde\rho_B+s_{A}+s'_A,\label{eq1addvv} \\
\frac{\partial \tilde\rho_{B}}{\partial t} &= D_B \nabla ^2 \tilde\rho_{B} 
-\msKappa\tilde\rho_A \tilde\rho_B+s_{B}+s'_B. \label{eq2ddavv}
\end{align}
These equations can be re-expressed in the standard form
\begin{align}
\frac{\partial \tilde\rho_{A}}{\partial t} -D_A \nabla ^2\tilde\rho_{A}  &= f_A(\vec{r}, t, \tilde\rho_{A}, \tilde\rho_{B})\\
\frac{\partial \tilde\rho_{B}}{\partial t} -D_B \nabla ^2\tilde\rho_{A}  &= f_B(\vec{r}, t, \tilde\rho_{A}, \tilde\rho_{B})
\end{align}
for $$f_A(\vec{r}, t, \tilde\rho_{A}, \tilde\rho_{B})=-\msKappa\tilde\rho_A \tilde\rho_B+s_{A}+s'_A,$$
$$f_B(\vec{r}, t, \tilde\rho_{A}, \tilde\rho_{B})=-\msKappa\tilde\rho_A \tilde\rho_B+s_{B}+s'_B.$$
Observe that 
$$f_A(\vec{r}, t, \tilde\rho_{A}, \tilde\rho_{B})-
f_A(\vec{r}, t, \tilde\rho_{A}, \tilde\rho_{B}')
=-\msKappa\tilde\rho_A (\tilde\rho_B-\tilde\rho_{B}'),$$
and $f_A$ and $f_B$ are quasi-monotone decreasing functions (in the sense of  \cite{Pao2}). As a result, \eqref{eq1avev232} and \eqref{eq1avev2}  imply that $(\rho_A+\rho'_A, \rho_B+\rho'_B)$ is an upper solution for \eqref{eq1addvv} and \eqref{eq2ddavv} in the sense of \cite{Pao2}. This completes the proof.

\end{document}